\documentclass[12pt, draftclsnofoot, onecolumn]{IEEEtran}
\usepackage{cite}
\usepackage{pgfplots} 
\usepackage[utf8]{inputenc}
\usepackage[english]{babel}
\usepackage{amsthm}
\usepackage{bm}
\usepackage{multicol}
\usepackage{multirow}
\usepackage[cmex10]{amsmath}
\usepackage{graphicx}
\usepackage{url}
\usepackage{color}
\usepackage{amsfonts}
\usepackage{amsmath}
\usepackage{extarrows}
\usepackage{amsfonts}
\usepackage{caption}
\usepackage{algorithm}
\usepackage{algpseudocode}
\usepackage{indentfirst}
\usepackage{caption}
\usepackage{esint} 
\usepackage{epsfig}
\usepackage{soul}
\usepackage{caption}
\usepackage{gensymb}
\usepackage{algpseudocode}  
\usepackage{amsmath} 
\usepackage{balance} 
\usepackage{algorithm,algpseudocode}

\makeatletter
\newcommand\fs@spaceruled{\def\@fs@cfont{\bfseries}\let\@fs@capt\floatc@ruled
  \def\@fs@pre{\vspace{1\baselineskip}\hrule height.8pt depth0pt \kern2pt}%
  \def\@fs@post{\kern2pt\hrule\relax}%
  \def\@fs@mid{\kern2pt\hrule\kern2pt}%
  \let\@fs@iftopcapt\iftrue}
\makeatother

\newtheorem{remark}{Remark}

\def\bA{{\mathbf{A}}}  \def\bC{{\mathbf{C}}} \def\bD{{\mathbf{D}}} \def\bE{{\mathbf{E}}}
 \def\bG{{\mathbf{G}}} \def\bH{{\mathbf{H}}} \def\bI{{\mathbf{I}}} \def\bJ{{\mathbf{J}}}

\def\bU{{\mathbf{U}}} \def\bV{{\mathbf{V}}} \def\bW{{\mathbf{W}}} \def\bX{{\mathbf{X}}} \def\bY{{\mathbf{Y}}}
\def\bZ{{\mathbf{Z}}}
\def\ba{{\mathbf{a}}} \def\bb{{\mathbf{b}}} \def\bc{{\mathbf{c}}}  
\def\bf{{\mathbf{f}}} \def\bg{{\mathbf{g}}} \def\bh{{\mathbf{h}}}  
   \def\bn{{\mathbf{n}}} 
    
\def\bu{{\mathbf{u}}} \def\bv{{\mathbf{v}}} \def\bw{{\mathbf{w}}}  \def\by{{\mathbf{y}}}
\def\bz{{\mathbf{z}}}
\def\tcb{\textcolor{black}}

\begin{document}
\captionsetup[figure]{name={Fig.},labelsep=period}

\title{Channel Estimation for RIS-Aided mmWave MIMO Systems via Atomic Norm Minimization
\thanks{J. He and M. Juntti are with Centre for Wireless Communications, FI-90014, University of Oulu, Finland.}
\thanks{H. Wymeersch is with Department of Electrical Engineering, Chalmers University of Technology, Gothenburg, Sweden.}
\thanks{This work is supported by Horizon 2020, European Union's Framework Programme for Research and Innovation, under grant agreement no. 871464 (ARIADNE). This work is also partially supported by the Academy of Finland 6Genesis Flagship (grant 318927) and Swedish Research Council (grant no. 2018-03701).}
}
\author{Jiguang~He,~\IEEEmembership{Member,~IEEE,} Henk~Wymeersch,~\IEEEmembership{Senior Member,~IEEE,} and Markku~Juntti,~\IEEEmembership{Fellow,~IEEE} }

\maketitle
\begin{abstract}
A reconfigurable intelligent surface (RIS) can shape the radio propagation environment by virtue of changing the impinging electromagnetic waves towards any desired directions, thus, breaking the general Snell's reflection law. However, the optimal control of the RIS requires perfect channel state information (CSI) of the individual channels that link the base station (BS) and the mobile station (MS) to each other via the RIS. Thereby super-resolution channel (parameter) estimation needs to be efficiently conducted at the BS or MS with CSI feedback to the RIS controller. \tcb{ In this paper, we adopt a two-stage channel estimation scheme for RIS-aided millimeter wave (mmWave) MIMO systems without a direct BS-MS channel, using atomic norm minimization to sequentially estimate the channel parameters, i.e., angular parameters, angle differences, and the products of propagation path gains.} We evaluate the mean square error of the parameter estimates, the RIS gains, the average effective spectrum efficiency bound, and average squared distance between the designed beamforming and combining vectors and the optimal ones. The results demonstrate that the proposed scheme achieves super-resolution estimation compared to the existing benchmark schemes, thus offering promising performance in the subsequent data transmission phase. 
\end{abstract}

\begin{IEEEkeywords}
Atomic norm minimization, channel parameter estimation, compressive sensing, millimeter wave MIMO, reconfigurable intelligent surface. 
\end{IEEEkeywords}

\section{Introduction}
The millimeter wave (mmWave) bands with multiple-input multiple-output (MIMO) transmission is a promising candidate for 5G and beyond 5G communication systems~\cite{Rappaport2013}. However, the transmission distance is limited due to the high free-space path loss, which can be compensated for by introducing large antenna arrays at both ends of the link~\cite{Alkhateeb2014,He2014,Heath2016}. This in turn brings challenges on the channel estimation (CE) compared to that for small-scale MIMO systems with less unknown channel coefficients. Unlike the sub-6 GHz bands, the wireless channels at mmWave frequencies are verified to have less scattering~\cite{Rappaport2013}. Thereby fewer resolvable paths exist between the base station (BS) and mobile station (MS). Thus, the mmWave MIMO channel is typically inherently sparse (i.e., the number of distinguishable paths in the angular domain is much smaller than that of transmit and receive antennas). Efficient yet effective compressive sensing (CS) techniques, which take advantage of the sparsity, have been widely applied in the channel (parameter) estimation of point-to-point (P2P) mmWave MIMO channels, e.g., in~\cite{Marzi2016, Lee2016,Hu2018,Tsai2018}. 

Due to the channel sparsity, the mmWave communications typically require line-of-sight (LoS) connection to maintain sufficient receive power level. In practice, the direct channel between the BS and MS can be blocked by objects~\cite{Raghavan2019}. In order to maintain the connectivity under LoS blockage, the concept of a reconfigurable intelligent surface (RIS), also known as intelligent reflecting surface (IRS)~\cite{Wu2019} or large intelligent surface (LIS)~\cite{Hu2018_SP1, Hu2018_SP}, has been recently proposed in~\cite{Liaskos18,2019Basar,Huang2018,di_renzo_smart_2019,He2019large} as a smart reflector. It can also been interpreted as a full-duplex (FD) relay~\cite{bjornson_intelligent_2019}, although it is in reality a passive element with no active transmit power amplifier, which is a core component of an actual relay station. Other potential benefits brought by introducing a RIS include enhanced spectrum efficiency (SE), energy efficiency (EE), and physical-layer security~\cite{Cui2019}, \tcb{which makes RIS a promising candidate for upcoming 6G~\cite{Rajatheva2020}}. Additionally, the RIS has potential to offer higher-accurate indoor or outdoor radio localization~\cite{He2019large,wymeersch2019radio}. In practice, the RIS can be made of an array of discrete phase shifters, which can passively steer  beams towards dedicated terminals by controlling the phase of each RIS unit. This kind of RIS architecture is called the \emph{discrete RIS} and does not have any baseband processing capability~\cite{2019Basar,Huang2018,He2019large}. Therefore, extremely low power consumption is expected, used only for the control of the RIS units. Another type of RIS, on the contrary, is the \emph{continuous/contiguous RIS}, which can be seen as an active transceiver with baseband processing capability~\cite{Hu2018_SP} or a passive reflector~\cite{huang2019holographic} like the aforementioned discrete RIS. \tcb{Various works on RIS channel modelling were conducted~\cite{ozdogan2020,Garcia2020,Najafi2020}, and these will guide the development of CE algorithm and design of RIS phase control matrix, studied in this paper. In these works, the RIS elements are modelled as individual scatterers and can be jointly considered for the purpose of steering the signal in a dedicated direction. Dynamic metasurface antennas with advanced analog signal processing capabilities for 6G communication were discussed in~\cite{Shlezinger2020} in terms of their main characteristics when used for radiation and reception. In addition, a hardware architecture with single radio frequency (RF) chain at the RIS was proposed explicitly for channel estimation purpose with alternating optimization method in~\cite{Alexandropoulos2020}. }

CE methods for RIS-aided MIMO systems have been recently studied in~\cite{Wu2020Mag,yuan2020,Taha2019,He2020,Jensen2020,Wang2020,he2020channel}. \tcb{The RIS channel estimation was discussed in~\cite{Wu2020Mag,yuan2020} as one of the main design challenges.} Taha \textit{et al.}~\cite{Taha2019} considered a special setup with mixed active and passive elements at the RIS. Therefore, CE was performed using CS and deep learning (DL) methods at the RIS based on the received signals at the active elements with pilots sent from the BS and MS. \tcb{The introduction of active receive elements at the RIS increases the power consumption, complexity and cost of RIS, but can simplify the CE problem into two P2P MIMO CE subproblems~\cite{schroeder2020passive}.} In~\cite{He2020}, sparse matrix factorization and matrix completion were exploited in a sequential manner to perform iterative CE. Thereby full rate advantage of the RIS is not achieved during the training process due to the on/off state applied to the RIS elements. \tcb{The individual MIMO channels in the reflection link can also be estimated by parallel factor decomposition~\cite{Wei2020,deAraujo2020}. In these works, iterative refinement of the individual channel estimation is conducted by using bilinear alternating least squares (BALS).} An optimal CE scheme was studied by following the criterion of minimum variance unbiased (MVU) estimation in~\cite{Jensen2020}. In~\cite{Wang2020},  CS was applied to estimate the cascade mmWave channel. However, a single antenna was assumed for the MS in both~\cite{Jensen2020} and~\cite{Wang2020}, which applies for wireless sensor network applications, but is not practical for mmWave MIMO communications. In our recent work~\cite{he2020channel}, we applied the iterative reweighted method of~\cite{Fang2016,Hu2018} to estimate the channel parameters. However, both BS-RIS and RIS-MS channels were assumed to have only a LoS path. Unlike all the aforementioned literature, a multi-level hierarchical codebook based scheme was leveraged to design the phase control matrix (reflection beam) at the RIS and the combining vector at the MS jointly~\cite{he2019adaptive} instead of estimating the MIMO channel parameters as an intermediate step towards joint design of active combining vector at the MS and passive beamforming (BF) at the RIS.

In this paper, we study the CE problem of passive RIS-aided mmWave MIMO systems, where the direct channel is obstructed and multiple paths exist for both the BS-RIS and RIS-MS channels. We resort to the parametric channel model for the individual channels~\cite{Alkhateeb2014,Ayach2014}, based on angular parameters, i.e., angles of departure (AoDs) and angles of arrival (AoAs), and propagation path gains. Furthermore, no data sharing backhaul link is assumed between the BS and RIS; low rate control link is sufficient. We divide the CE problem into two CS subproblems and apply atomic norm minimization to sequentially find the estimates of the channel parameters, e.g., angular parameters, angle differences, and the products of propagation path gains.  \tcb{We take advantage of channel sparsity in the proposed CE algorithm. Unlike the estimation of cascaded channel or individual channels, much fewer elements need to be estimated. In addition, when the number of elements (including both RIS elements and BS/MS antennas) increases, estimation of individual channel matrices or cascaded channel matrix will cause substantial increases in both training overhead and computational complexity. On the contrary, channel sparsity level will further increase, which may even reduce the required training overhead.} 
Besides evaluating the mean square error (MSE) of the estimated channel parameters, we design the RIS phase control matrix, the BS BF vector, and the MS combining vector based on the estimates and evaluate the average effective SE bound and RIS gains. The proposed CE scheme significantly outperforms an orthogonal matching pursuit (OMP) based two-stage counterpart~\cite{Tropp2007}. Simulation results demonstrate that the average effective SE bound achieved by the proposed method approximate that with perfect channel state information (CSI) in the low signal-to-noise ratio (SNR) regime with limited training overhead. 
The contributions of the paper are summarized as follows:
\begin{itemize}
\item We propose an efficient super-resolution channel parameter estimation scheme for RIS-aided mmwave MIMO systems, based on atomic norm minimization~\cite{Tang2013,ZHANG201995}. The proposed scheme can reduce the training overhead significantly by first estimating part of the channel parameters (i.e., AoDs of the BS-RIS channel and AoAs of the RIS-MS channel) and utilizing the estimates in the subsequent training period.  

\item Decoupled atomic norm minimization is applied in the first stage with a multiple measurement vectors (MMV) model \tcb{for the estimation of AoDs of the BS-RIS channel and AoAs of the RIS-MS channel}, while atomic norm minimization is applied in the second stage with a single measurement vector (SMV) one \tcb{for the estimation of angle differences and the products of propagation path gains}. 

\item The design of RIS phase control matrix is studied by following the criterion of maximizing the power of the effective channel. On the basis of the designed RIS phase control matrix, the joint design of BS BF and MS combining vectors are considered based on the reconstructed composite channel matrix (using estimated channel parameters). 
\end{itemize}

The rest of the paper is organized as follows: Section~\ref{section_channel_model} introduces the channel model for the RIS-aided mmWave MIMO system, followed by the sounding procedure in Section~\ref{section_sounding_procedure}. Section~\ref{section_two_stage_CE_approach} provides the details about the proposed two-stage CE approach based on atomic norm minimization, followed by the RIS control as well as beamforming and combining design in Section \ref{sec:Control}. The performance evaluation is offered in Section~\ref{section_performance_evaluation}. Section~\ref{section_conclusion_future_work} draws the conclusions and discusses the potential directions for future investigation. 

\textit{Notations:} A bold lowercase letter $\ba$ denotes the column vector, a bold capital letter $\bA$ denotes the matrix, $(\cdot)^{\mathsf{H}}$, $(\cdot)^{\mathsf{T}}$, and $(\cdot)^*$ denote the Hermitian transpose, transpose, and conjugate, respectively, $\mathrm{diag}(\ba)$ denotes a square diagonal matrix with entries of $\ba$ on its diagonal, $\mathrm{Toep}(\ba)$ is a Toeplitz matrix with $\ba$ being its first row, $\mathrm{Tr}(\bA)$ returns the sum value of the diagonal elements of $\bA$, $\mathrm{vec}(\bA)$ denotes the vectorization of $\bA$ by stacking the columns of the matrix $\bA$ on top of one another, $\mathbb{E}[\cdot]$ is the expectation operator, $\mathrm{var}(\cdot)$ is the variance of a random variable, $\lceil a \rceil$ returns the least integer greater than or equal to $a$, $\ba \circ \bb$ and $\ba \otimes \bb$ denote the Hadamard product and Kronecker product of $\ba$ and $\bb$, respectively, $[\ba]_i$ denotes the $i$th element of vector $\ba$, $[\bA]_{ij}$ denotes the $(i,j)$th element of~$\bA$, $[\bA]_{i,:}$ and $[\bA]_{:,i}$ denote the $i$th row and column vectors of $\bA$, respectively, $\bA \succeq \mathbf{0}$ means $\bA$ is positive semidefinite, and $\|\cdot\|_{\mathrm{F}}$ is the Frobenius norm.


\section{Channel Model}\label{section_channel_model}
\begin{figure}[t]
	\centering
	\includegraphics[width=0.65\linewidth]{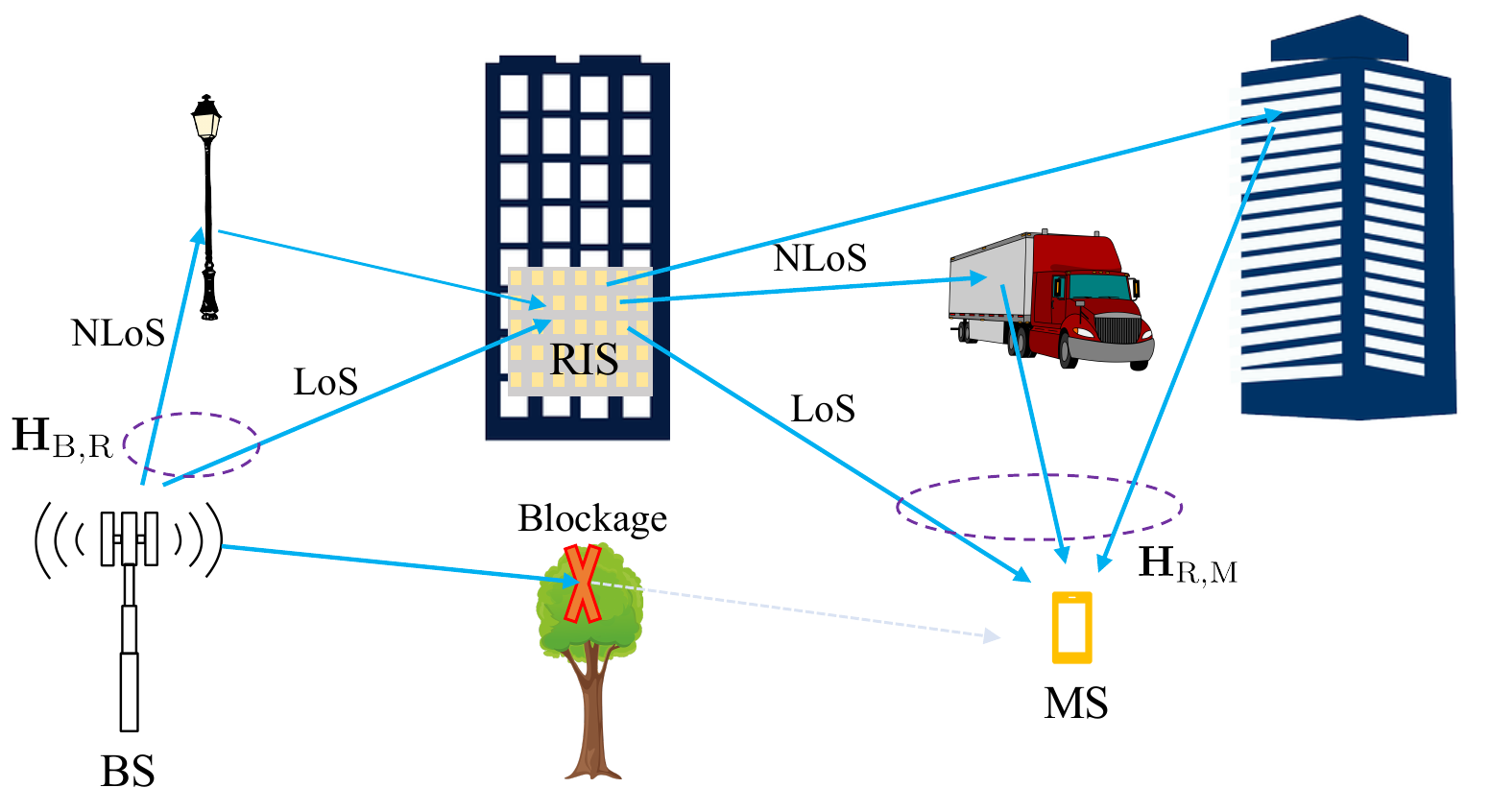}
	\caption{The considered RIS-aided mmWave MIMO system with one multi-antenna BS, one multi-antenna MS, and one multi-element RIS, with $L_{\text{B,R}}=2$ resolvable paths between the BS and RIS and $L_{\text{R,M}}=3$ resolvable paths between the RIS and MS.}
	\label{System_model}
\end{figure}
We consider the RIS-aided mmWave MIMO system, which comprises one multi-antenna BS, one multi-antenna MS, and one multi-element RIS, as depicted in Fig.~\ref{System_model}. No data-sharing backhaul link is assumed between the BS and RIS. The numbers of antenna elements at BS and MS are denoted as $N_{\text{B}}$ and $N_{\text{M}}$, respectively; the number of elements at the RIS is $N_{\text{R}}$. The antenna array is assumed to be an uniform linear array (ULA) with consideration of azimuth angle only; an extension to an uniform planar array (UPA) can be done.\footnote{Fig.~\ref{System_model} shows the RIS as an UPA for the sake of better aesthetic illustration. The proposed channel estimation scheme can also be extended to an UPA-type RIS-aided mmWave MIMO system with some modifications.} We further assume that the direct channel between the BS and MS is obstructed, which renders the potential usage of a RIS for maintaining the connectivity between the BS and MS.\footnote{The proposed scheme can also be applied to the scenario, where the direct BS-MS channel also exists. The process is summarized as follows: In the first step, we turn the RIS into an absorption mode, and estimate the direct channel, i.e., BS-MS channel; In the second step, we apply the proposed scheme to estimate the channel parameters in the composite channel, i.e., BS-RIS-MS channel.}

We assume the geometric channel model, which is based on the AoDs, the AoAs, and the propagation path gains of each link. \tcb{The channel model was also validated in the recent works~\cite{ozdogan2020,Garcia2020,Najafi2020}.} The channel between the BS and the RIS $\bH_{\text{B,R}} \in \mathbb{C}^{N_{\text{R}} \times N_{\text{B}}}$ is 
\begin{align}\label{H_BS_RIS}
\bH_{\text{B,R}} &= \sum\limits_{l = 1}^{L_{\text{B,R}}} [\boldsymbol{\rho}_{\text{B,R}}]_l \boldsymbol{ \alpha}([\boldsymbol{\phi}_{\text{B,R}}]_l ) \boldsymbol{\alpha}^{\mathsf{H}}([\boldsymbol{\theta}_{\text{B,R}}]_l)\nonumber\\
 &= \bA(\boldsymbol{\phi}_\text{B,R}) \mathrm{diag}(\boldsymbol{\rho}_{\text{B,R}}) \bA^{\mathsf{H}}(\boldsymbol{\theta}_\text{B,R}),
\end{align}
where $[\boldsymbol{\theta}_{\text{B,R}}]_l$ and $[\boldsymbol{\phi}_{\text{B,R}}]_l$ denote the $l$th AoD and AoA of the BS-RIS channel, respectively, $L_{\text{B,R}}$ denotes the number of resolvable paths, which is usually on the order of 3--8 in mmWave frequency bands~\cite{Rappaport2013}, and $[\boldsymbol{\rho}_{\text{B,R}}]_l$ denotes the $l$th propagation path gain. Index ${l=1}$ refers to the LoS path, and $l>1$ refer to the non-line-of-sight (NLoS) paths, e.g., single-bounce or multi-bounce reflection paths. Usually, $|[\boldsymbol{\rho}_{\text{B,R}}]_1|^2 \gg |[\boldsymbol{\rho}_{\text{B,R}}]_l|^2$ for $l>1$, and the difference is easily more than 20 dB~\cite{Akdeniz2014}. Finally, ${\boldsymbol{ \alpha}([\boldsymbol{\theta}_{\text{B,R}}]_l ) \in \mathbb{C}^{N_{\text{B}} \times 1}}$ and ${\boldsymbol{\alpha}([\boldsymbol{\phi}_{\text{B,R}}]_l)\in \mathbb{C}^{N_{\text{R}} \times 1}}$ are the array response vectors with $\big[\boldsymbol{ \alpha}([\boldsymbol{\theta}_{\text{B,R}}]_l )\big]_k = \exp\big(j 2 \pi \frac{d}{\lambda} (k-1) \sin([\boldsymbol{\theta}_{\text{B,R}}]_l)\big)$ for $k =1,\cdots, N_{\text{B}}$ and $\big[\boldsymbol{\alpha}([\boldsymbol{\phi}_{\text{B,R}}]_l)\big]_k = \exp\big(j 2 \pi \frac{d}{\lambda} (k-1) \sin([\boldsymbol{\phi}_{\text{B,R}}]_l)\big)$ for $k =1,\cdots, N_{\text{R}}$, where $d$ is the antenna element spacing, $\lambda$ is the wavelength of the carrier frequency, and $j \overset{\triangle}{=} \sqrt{-1}$. By following $\boldsymbol{\phi}_\text{B,R} = \big[[\boldsymbol{\phi}_{\text{B,R}}]_1, \cdots,[\boldsymbol{\phi}_{\text{B,R}}]_{L_{\text{B,R}}}\big]^{\mathsf{T}}$ and $\boldsymbol{\theta}_\text{B,R} = \big[[\boldsymbol{\theta}_{\text{B,R}}]_1, \cdots, [\boldsymbol{\theta}_{\text{B,R}}]_{L_{\text{B,R}}}\big]^{\mathsf{T}}$, array response matrices $\bA(\boldsymbol{\phi}_\text{B,R}) \in \mathbb{C}^{N_{\text{R}} \times L_{\text{B,R}}}$ and $\bA(\boldsymbol{\theta}_\text{B,R})\in \mathbb{C}^{N_{\text{B}} \times L_{\text{B,R}}}$ are formulated as 
\begin{align}
\bA(\boldsymbol{\theta}_\text{B,R}) & =\Big[ \boldsymbol{ \alpha}\big([\boldsymbol{\theta}_{\text{B,R}}]_1), \cdots,  \boldsymbol{\alpha}([\boldsymbol{\theta}_{\text{B,R}}]_{L_{\text{B,R}}} \big)\Big], \\
\bA(\boldsymbol{\phi}_\text{B,R}) & = \Big[ \boldsymbol{ \alpha}\big([\boldsymbol{\phi}_{\text{B,R}}]_1 ), \cdots,  \boldsymbol{ \alpha}([\boldsymbol{\phi}_{\text{B,R}}]_{L_{\text{B,R}}} \big)\Big].
\end{align}

Similar to~\eqref{H_BS_RIS}, the channel between the RIS and the MS, denoted as $\bH_{\text{R,M}} \in \mathbb{C}^{N_{\text{M}} \times N_{\text{R}}}$, is 
\begin{align}\label{H_RIS_MS}
\bH_{\text{R,M}} &= \sum\limits_{l = 1}^{L_{\text{R,M}}} [\boldsymbol{\rho}_{\text{R,M}}]_l \boldsymbol {\alpha}([\boldsymbol{\phi}_{\text{R,M}}]_l ) \boldsymbol{\alpha}^{\mathsf{H}}([\boldsymbol{\theta}_{\text{R,M}}]_l )\nonumber\\
&= \bA(\boldsymbol{\phi}_\text{R,M}) \mathrm{diag}(\boldsymbol{\rho}_{\text{R,M}}) \bA^{\mathsf{H}}(\boldsymbol{\theta}_\text{R,M}),
\end{align}
where the channel parameters $\boldsymbol{\phi}_\text{R,M}$, $\boldsymbol{\rho}_{\text{R,M}}$, $\boldsymbol{\theta}_\text{R,M}$, $\bA(\boldsymbol{\phi}_\text{R,M})$, and $\bA(\boldsymbol{\theta}_\text{R,M})$ are defined in the same manner as those in~\eqref{H_BS_RIS}.

Using~\eqref{H_BS_RIS} and~\eqref{H_RIS_MS}, the composite channel ${\bH \in \mathbb{C}^{N_{\text{M}} \times N_{\text{B}}}}$ between the BS and MS, after taking into consideration the RIS, becomes 
\begin{align}\label{Entire_channel}
\bH & =  \bH_{\text{R,M}} \boldsymbol{\Omega} \bH_{\text{B,R}}\nonumber\\
& = \bA(\boldsymbol{\phi}_\text{R,M}) \mathrm{diag}(\boldsymbol{\rho}_{\text{R,M}}) \bA^{\mathsf{H}}(\boldsymbol{\theta}_\text{R,M}) \nonumber\\
& \phantom{nn} \boldsymbol{\Omega}\bA(\boldsymbol{\phi}_\text{B,R}) \mathrm{diag}(\boldsymbol{\rho}_{\text{B,R}}) \bA^{\mathsf{H}}(\boldsymbol{\theta}_\text{B,R}),
\end{align}
where $\boldsymbol{\Omega}\in\mathbb{C}^{N_{\text{R}}\times N_{\text{R}}}$ is the phase control matrix at the RIS. We assume that the RIS is composed of a series of discrete phase shifters. Therefore, matrix $\boldsymbol{\Omega}$ is a diagonal matrix with unit-modulus constraint on the diagonal entries, i.e., $[\boldsymbol{\Omega}]_{kk} = \exp(j \omega)$ with phase $\omega \in [0,\;\; 2\pi)$. In practice, the reflection of RIS may not be perfect so that reflection coefficient $a \in [0,\;1]$ as in $[\boldsymbol{\Omega}]_{kk} = a \exp(j \omega)$ describes the amplitude scaling and power loss\footnote{If $a = 0$, the RIS is assumed to be operating in an absorption mode. On the contrary, if $a = 1$, the RIS is assumed to be operating in an ideal reflection mode. In practice, due to the imperfect fabrication of RIS elements, the reflection coefficients may vary from one RIS element to another.}~\cite{Wu2019}. We assume an ideal RIS with $a = 1$; for our focus on CE, this does not decrease the generality of the work as long as the value of $a$ is known.\footnote{\tcb{However, in practice, phase-dependent amplitude variation may exist in  the  RIS elements~\cite{Abeywickrama2020}, which may require redesign of the proposed CE scheme and RIS phase control matrix.}} In this regard, the received power at the MS can be considered as a theoretical upper bound if the RIS phase control matrix is optimally designed. 

Let us define $\bG \in \mathbb{C}^{L_{\text{R,M}} \times L_{\text{B,R}}}$ as the effective channel, 
\begin{equation}\label{G_matrix}
\bG =  \mathrm{diag}(\boldsymbol{\rho}_{\text{R,M}}) \bA^{\mathsf{H}}(\boldsymbol{\theta}_\text{R,M}) \boldsymbol{\Omega}\bA(\boldsymbol{\phi}_\text{B,R}) \mathrm{diag}(\boldsymbol{\rho}_{\text{B,R}}),
\end{equation}
taking into consideration of propagation path gains, RIS phase control matrix and the angular parameters associated with the RIS, i.e., $\boldsymbol{\theta}_\text{R,M}$ and $\boldsymbol{\phi}_\text{B,R}$. \tcb{The expression~\eqref{G_matrix} will be utilized in the second CE stage, discussed in Section~\ref{subsect:2nd_state} and the design of phase control matrix based on parameter estimates, discussed in Section~\ref{sec:Control}.} Because $\bG$ is a function of the RIS phase control matrix, the design of $\boldsymbol{\Omega}$ affects the effective channel, which in turn influences the achievable rate (i.e., capacity) of the composite channel. This imposes the significance of the RIS design and control for data communications, especially, when the direct BS-MS channel is blocked. By following~\eqref{G_matrix}, the composite channel $\bH$ in~\eqref{Entire_channel} can be further expressed as 
\begin{equation}\label{Channel_matrix}
\bH  =  \bA(\boldsymbol{\phi}_\text{R,M}) \bG \bA^{\mathsf{H}}(\boldsymbol{\theta}_\text{B,R}).
\end{equation}
\begin{remark}
The composite channel matrix $\bH$ in~\eqref{Channel_matrix} is similar to a P2P mmWave MIMO channel. However, a difference exists. As for the P2P mmWave MIMO channel, $\bG$ is a diagonal matrix, like $\mathrm{diag}(\boldsymbol{\rho}_{\text{B,R}}) $ in~\eqref{H_BS_RIS} and $\mathrm{diag}(\boldsymbol{\rho}_{\text{R,M}}) $ in~\eqref{H_RIS_MS} while for the RIS-aided MIMO channel, $\bG$ is usually in a general format, i.e., a full matrix. In addition, the effective channel matrix $\bG$ needs to be optimized via controlling the RIS phase shifters in order to take the full potential of introducing the RIS. 
\end{remark}


\tcb{In the first CE stage, we estimate $\boldsymbol{\phi}_\text{R,M}$ and $\boldsymbol{\theta}_\text{B,R}$ with randomly generated training sequences. In the second CE stage, we estimate the remaining channel parameters, e.g., $\boldsymbol{\rho}_{\text{R,M}}$, $\boldsymbol{\theta}_\text{R,M}$, $\boldsymbol{\rho}_{\text{B,R}}$, and $\boldsymbol{\phi}_\text{B,R}$ based on the training sequences designed according to the estimates in the first stage. Due to the coupling effect in~\eqref{G_matrix}, these parameters cannot be estimated separately in the second stage, detailed in Section~\ref{section_two_stage_CE_approach}.}

\section{Sounding Procedure}\label{section_sounding_procedure}
\begin{figure}[t]
	\centering
	\includegraphics[width=0.65\linewidth]{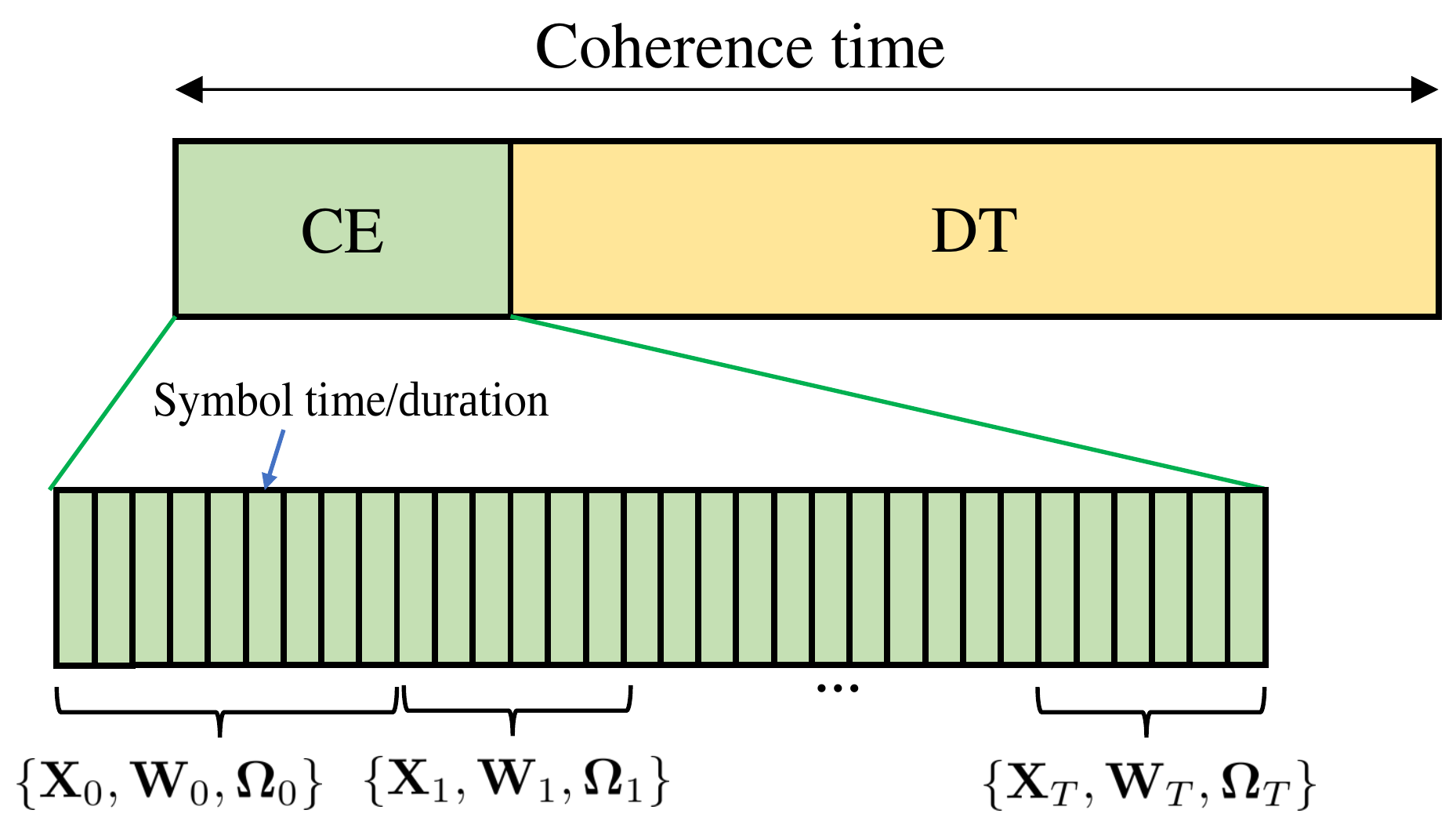}
	\caption{The sounding procedure, where each CE subinterval contains $T+1$ blocks (indexed by $t = 0,\cdots, T$) and $\boldsymbol{\Omega}_t$ varies over the blocks. \tcb{In the example, the phase control matrix keeps unchanged within the first $9$ symbol times (i.e., the first block of the CE subinterval, also known as stage 1 sounding), and varies every $6$ symbol times in the stage 2 sounding.}}
	\label{Sounding_procedure}
\end{figure}

 We also assume that the wireless channels are quasi-static block fading. That is, the channel parameters remain unchanged during a certain period of time, known as the coherence time. For the sounding process, one coherence time interval is divided into two subintervals, the first one for CE and the second for data transmission (DT), as depicted in Fig.~\ref{Sounding_procedure}. The CE subinterval is further divided into $T+1$ blocks. In each block, a different $\boldsymbol{\Omega}$ is taken into consideration, i.e., $\boldsymbol{\Omega}_0 \neq \boldsymbol{\Omega}_1 \neq \cdots \neq \boldsymbol{\Omega}_{T}$. \tcb{The frequent change of the RIS phase control matrix within one coherence time can be achieved by n-type field-effect transistor (nFET) switches. The turn-on and turn-off times of the switch are on the order of 300 ps~\cite{Schmid2014}, which can be much smaller than a symbol duration at mmWave communications.}  

\subsection{Stage 1 Sounding}
In the first block of CE subinterval, i.e., $t=0$, the BS sends a (random) training matrix $\bX_0 \in \mathbb{C}^{N_{\text{B}}\times N_{0}}$ which, after reflected from the RIS with a (random) phase control matrix $\boldsymbol{\Omega}_0$,\footnote{The phase control matrix is assumed to be known to the MS. This can be achieved by generating it by agreed pseudo-noise (PN) sequences.} is received at the MS as $\bY_0\in \mathbb{C}^{ M_0 \times N_0}$ through a (random) combining matrix $\bW_0 \in \mathbb{C}^{N_{\text{M}}\times M_0}$. As in mmWave MIMO systems, the BS and MS are commonly assumed to possess a hybrid analog-digital precoding architecture with limited number of RF chains for the sake of reduced complexity, cost, and power consumption~\cite{Alkhateeb2014,Hu2018TVT,Ayach2014,He2014}. We follow the same hybrid architecture in this paper. Therefore, at the MS, we can only access to a maximum $N_{\text{RF}}$-dimensional signal vector per symbol time\footnote{\tcb{The coherence time interval may include hundreds or even thousands of modulated symbol times/durations at mmWave frequency bands, e.g., in~\cite{Yang2018}, which depends on the carrier frequency, MS velocity, and the bandwidth.}} with $N_{\text{RF}}$ being the number of RF chains at the MS. In other words, the combining matrix at the MS can be as large as $N_{\text{M}}\times N_{\text{RF}}$ per symbol duration. Meanwhile, at the BS, we can only explore one beam (i.e., one column vector of transmitted signals in $\bX_0$) per symbol duration regardless of the number of RF chains at the BS~\cite{Alkhateeb2014,Hu2018TVT}. When $N_{\text{RF}} < M_0$, each training beam from $\bX_0$ needs to be sent $\lceil \frac{M_0}{N_{\text{RF}}}\rceil$ times. Thus, the training overhead in the first stage is $N_0 \lceil \frac{M_0}{N_{\text{RF}}}\rceil$~\cite{Hu2018}. 

\subsection{Stage 2 Sounding}
Based on the received signal $\bY_0$, we resort to the atomic norm minimization to recover the angular parameters $\boldsymbol{\theta}_\text{B,R}$ and $\boldsymbol{\phi}_\text{R,M}$, which guide the design of sequential training matrices $\{\bX_1, \cdots, \bX_T\}$ and combining matrices $\{\bW_1, \cdots, \bW_T\}$. To simplify the design, we fix $\bX_1 = \cdots = \bX_T \in \mathbb{C}^{{N_\text{B}} \times L_\text{B,R}}$ and $\bW_1= \cdots=\bW_T\in \mathbb{C}^{{N_\text{M}} \times L_\text{R,M}}$ while changing $\boldsymbol{\Omega}_t$ for $t = 1, \cdots, T$ and obtain the received signals as $\{\bY_1, \cdots, \bY_T\}$.\footnote{In principle, we can refine the training and combing matrices at block $t$ based on the received signals up to block $t-1$. However, this will bring more computational complexity of the proposed CE algorithm. Also, we intentionally use more time slots in the first block of CE subinterval in order to obtain a super resolution for the estimates of channel parameters in the first stage. Therefore, the room for gradual improvement will be rather limited. } We intentionally choose $N_0 \gg L_\text{B,R}$ and $M_0 \gg L_\text{R,M}$ in order to provide a very accurate estimate in the first stage. Therefore, the training overhead can be greatly reduced for the block $t$ as $t = 1, \cdots, T$ compared to that for the first block. The overall training overhead in the second stage is $T L_{\text{B,R}} \lceil \frac{L_{\text{R,M}}}{N_{\text{RF}}}\rceil$. Based on $\{\bY_1, \cdots, \bY_T\}$, the atomic norm minimization is further applied to estimate the remaining channel parameters as detailed below.

\subsection{Observation Model}
The received signals for all the blocks are summarized as 
\begin{align}\label{Rec_signal1}
\bY_t &= \bW_t^{\mathsf{H}} \bH(\boldsymbol{\Omega}_t) \bX_t + \bW_t^{\mathsf{H}} \bZ_t, \nonumber\\
&=  \bW_t^{\mathsf{H}} \bA(\boldsymbol{\phi}_\text{R,M}) \bG_t \bA^{\mathsf{H}}(\boldsymbol{\theta}_\text{B,R})\bX_t + \bW_t^{\mathsf{H}} \bZ_t,
\nonumber\\
& \;\;\;\;\;\;\text{for}\; t =0, \cdots, T,
\end{align}
where we write $\bH$ explicitly as a function of $\boldsymbol{\Omega}_t$, $\bG_t =  \mathrm{diag}(\boldsymbol{\rho}_{\text{R,M}}) \bA^{\mathsf{H}}(\boldsymbol{\theta}_\text{R,M}) \boldsymbol{\Omega}_t\bA(\boldsymbol{\phi}_\text{B,R}) \mathrm{diag}(\boldsymbol{\rho}_{\text{B,R}})$, and each entry in additive white Gaussian noise (AWGN) $\bZ_t$ follows $\mathcal{CN}(0,\sigma^2)$.

\section{Two-Stage CE Approach}\label{section_two_stage_CE_approach}
Before moving to the details of the two-stage CE approach, we briefly review the atomic set, the atomic norm, and the atomic norm minimization. 

\subsection{Atomic Norm Minimization}
Unlike the conventional greedy CS approaches, e.g., OMP, the atomic norm minimization is based on an infinite set and solved by resorting to convex optimization tools~\cite{Tang2013,yang2016}. Atomic norm minimization can well address the basis mismatch problem, which is commonly known in finite-size dictionary based CS approaches. Depending on the signals to be recovered, an atomic set is formulated by containing atoms with the same dimension of the desired signals~\cite{Tang2013,yang2016}.
\subsubsection{1D Signal}
As in direction of arrival (DoA) estimation or line spectral estimation problems~\cite{Bhaskar2013,Tang2013}, the one dimensional (1D) signal to be recovered is in the form of $\boldsymbol{\alpha}(\theta)\in \mathbb{C}^{N_u \times 1}$.\footnote{The ultimate goal is to recover the angle (e.g., DoA $\theta$) or equivalently frequency (e.g., $f = \sin(\theta)$), which is contained in vector $\boldsymbol{\alpha}(\theta)$ or equivalently in $\boldsymbol{\alpha}(f)$. Knowing $\boldsymbol{\alpha}(\theta)$ is tantamount to knowing $\theta$, and the same principle is applied to $\boldsymbol{\alpha}(f)$ and $f$, unless the following ambiguity exists, $\exists\;\boldsymbol{\alpha}(\theta_1) = \boldsymbol{\alpha}(\theta_2)$ with $\theta_1 \neq \theta_2$ or $\exists\;\boldsymbol{\alpha}(f_1) = \boldsymbol{\alpha}(f_2)$ with $f_1 \neq f_2$.} Therefore, the atomic set is defined as 
\begin{equation}\label{atomic_set1D}
\mathcal{A} = \{ \boldsymbol{ \alpha}(\theta_1)\in \mathbb{C}^{N_u \times 1}: \theta_1 \in [-\pi,\; \pi]\},
\end{equation}
where the cardinality of $\mathcal{A}$ is infinite, i.e., $\mathrm{card}(\mathcal{A} ) = +\infty$. For any signal with the same dimension of the atoms, e.g., $\bu\in \mathbb{C}^{N_u \times 1}$, its atomic norm with respect to $\mathcal{A}$ in~\eqref{atomic_set1D} is defined as 
\begin{align}
&\|\bu\|_{\mathcal{A}} =\mathrm{inf}\{q: \bu \in q \mathrm{conv} (\mathcal{A})\},\nonumber\\
&=\mathrm{inf}_{\{\theta_{1,l} \in [-\pi,\; \pi], \beta_l \in \mathbb{C}\}} \Big\{ \sum_{l} |\beta_l| \Big|\bu = \sum_{l} \beta_l \boldsymbol{ \alpha}(\theta_{1,l}) \Big\},
\end{align}
where $\mathrm{conv}(\mathcal{A})$ is the convex hull of $\mathcal{A}$, and $\bu = \bA_\text{u} \boldsymbol{\beta}$ falls into the SMV model with $\bA_\text{u} = [ \boldsymbol{ \alpha}(\theta_{1,1}), \boldsymbol{ \alpha}(\theta_{1,2}), \cdots]$ and $\boldsymbol{\beta} = [\beta_1, \beta_2, \cdots]^{\mathsf{T}}$. 

The atomic norm is equivalent to the solution of the following semidefinite program (SDP)~\cite{yang2016}
\begin{align}
\|\bu\|_{\mathcal{A}} = &\mathrm{inf}_{\{\bu_1, z\}}\Big\{\frac{z}{2}+ \frac{1}{2 N_{u}} \mathrm{Tr}(\mathrm{Toep}(\bu_1)) \Big\}, \nonumber\\
&\text{s.t.} \;\begin{bmatrix} \mathrm{Toep}(\bu_1)  & \bu\\
\bu^{\mathsf{H}}& z
\end{bmatrix} \succeq \mathbf{0}.
\end{align}

\subsubsection{2D Signal}
As for a two-dimensional signal, one valid matrix atomic set can be defined as~\cite{Tsai2018}
\begin{equation}\label{atomic_set2D}
\mathcal{A}_M = \{ \boldsymbol{ \alpha}(\theta_1) \bc^{\mathsf{T}} \in \mathbb{C}^{N_U \times M_U}: \theta_1 \in [-\pi,\; \pi], \|\bc\| = 1\}.
\end{equation}
We intentionally introduce such an atomic set, since it will be used in the first stage of the proposed two-stage CE scheme. Other types of matrix atomic sets also exist in the literature depending on the structure of the original signal to be recovered. Each atom in set $\mathcal{A}_M$ is a rank-1 matrix, and the atomic set size is also infinite due to the continuum of $\theta_1$.

For any matrix $\bU\in \mathbb{C}^{N_U \times M_U}$ with the same dimension of $\boldsymbol{ \alpha}(\theta_1)\bc^{\mathsf{T}}$, its atomic norm with respect to $\mathcal{A}_M$ in~\eqref{atomic_set2D} is defined as 
\begin{align}
&\|\bU\|_{\mathcal{A}_M} =\mathrm{inf}\{q: \bU \in q \mathrm{conv} (\mathcal{A}_M)\},\nonumber\\
&=\mathrm{inf}_{\{\theta_{1,l} \in [-\pi,\; \pi], \beta_l \in \mathbb{C}\}} \Big\{ \sum_{l} |\beta_l| \Big|\bU = \sum_{l} \beta_l \boldsymbol{ \alpha}(\theta_{1,l}) \bc_l^{\mathsf{T}} \Big\},
\end{align}
where $\mathrm{conv}(\mathcal{A}_M)$ is the convex hull of $\mathcal{A}_M$ and $\bU= \bA_\text{u} {\mathrm{diag}(\boldsymbol{\beta}) \bC^{\mathsf{T}} } = \bA_\text{u}\breve{\bC}$ falls into the MMV model with $\bC = [\bc_1, \bc_2, \cdots]$ and $\breve{\bC} = \mathrm{diag}(\boldsymbol{\beta}) \bC^{\mathsf{T}}$. This atomic norm is equivalent to the solution of the following SDP, as in~\cite{yang2016}
\begin{align}
\|\bU\|_{\mathcal{A}_M} = &\mathrm{inf}_{\{\bu_1, \bZ\}}\Big\{\frac{1}{2M_U}\mathrm{Tr}(\bZ) + \frac{1}{2 N_U} \mathrm{Tr}(\mathrm{Toep}(\bu_1))\Big\}, \nonumber\\
&\text{s.t.} \;\begin{bmatrix} \mathrm{Toep}(\bu_1)  & \bU\\
\bU^{\mathsf{H}}& \bZ
\end{bmatrix} \succeq \mathbf{0}.
\end{align}
Similar to other CS methods, the goal of atomic norm minimization is also to find the sparsest representation of $\bu$ or $\bU$ with the least number of atoms from the predefined atomic set~\cite{yang2016}.  

\subsection{First Stage of Channel Estimation Algorithm}
The CE problem in the first stage falls into the category of two decoupled 2D signal (with a MMV model) recovery subproblems. 

\subsubsection{Estimation of $\boldsymbol{\phi}_\text{R,M}$}
By expression $\bar{\bU} = \bA(\boldsymbol{\phi}_\text{R,M}) \bG_0 \bA^{\mathsf{H}}(\boldsymbol{\theta}_\text{B,R}) \bX_0$ as  $ \bar{\bU}=\bA(\boldsymbol{\phi}_\text{R,M}) \bar{\bC} $ with $\bar{\bC} = \bG_0 \bA_t^{\mathsf{H}}(\boldsymbol{\theta}_\text{B,R}) \bX_0$, the estimation of $\boldsymbol{\phi}_\text{R,M}$ based on $\bY_0$ in the first stage can be formulated as regularized denoising 
\begin{equation}
\min \frac{\mu}{2} \|\bar{\bU}\|_{\mathcal{A}_M} + \frac{1}{2}\|\bY_0-  \bW_0^{\mathsf{H}}\bar{\bU}\|_{\mathrm{F}}^2, 
\end{equation}
which can be further expressed as 
\begin{align}\label{ANM_Phi_first}
\{ \hat{\bar{\bu}}_1, \hat{\bar{\bZ}}, \hat{\bar{\bU}}\}=&\arg\min_{\bar{\bu}_1,\bar{\bZ},\bar{\bU}} \frac{\mu}{2 N_0}\mathrm{Tr}(\bar{\bZ}) + \frac{\mu}{2 N_{\text{M}}} \mathrm{Tr}(\mathrm{Toep}(\bar{\bu}_1))\nonumber\\
& + \frac{1}{2}\|\bY_0-  \bW_0^{\mathsf{H}}\bar{\bU}\|_{\mathrm{F}}^2 \nonumber\\
&\text{s.t.} \;\begin{bmatrix} \mathrm{Toep}(\bar{\bu}_1)  & \bar{\bU}\\
\bar{\bU}^{\mathsf{H}}& \bar{\bZ}
\end{bmatrix} \succeq \mathbf{0},
\end{align}
where $\mu$ is a regularization parameter controlling the trade-off between sparsity and data fitting, set as $\mu \propto \sqrt{\sigma^2 N_{\text{M}} \log(N_{\text{M}})}$~\cite{ZHANG201995}. We assume that we know the number of (significant) paths as prior information. In practice, this can be identified either by long-term site specific measurements or CS based support recovery algorithms, for example. The recovery of $\boldsymbol{\phi}_{\text{R,M}}$ is then based on the solution of $\mathrm{Toep}(\hat{\bar{\bu}}_1)$ from~\eqref{ANM_Phi_first} by root finding approach or other related approaches, e.g., the classical multiple signal classification (MUSIC) and estimation of signal parameters via rotational invariant techniques (ESPRIT)~\cite{Schmidt1986,Roy1989}. 

\subsubsection{Estimation of $\boldsymbol{\theta}_\text{B,R}$}
Similarly, based on the $\bY_0^{\mathsf{H}}$, we can recover $\boldsymbol{\theta}_\text{B,R}$ by addressing the following convex problem 
\begin{equation}
\min \frac{\eta}{2} \|\tilde{\bU}\|_{\mathcal{A}_M} + \frac{1}{2}\|\bY_0^{\mathsf{H}}-  \bX_0^{\mathsf{H}}\tilde{\bU}\|_{\mathrm{F}}^2, 
\end{equation}
where $\tilde{\bU} = \bA(\boldsymbol{\theta}_\text{B,R})  \bG_0^{\mathsf{H}} \bA^{\mathsf{H}}(\boldsymbol{\phi}_\text{R,M}) \bW_0 =  \bA(\boldsymbol{\theta}_\text{B,R}) \tilde{\bC}$ with $\tilde{\bC} =  \bG_0^{\mathsf{H}} \bA^{\mathsf{H}}(\boldsymbol{\phi}_\text{R,M}) \bW_0$, and $\eta$ is a regularization parameter controlling the trade-off between sparsity and data fitting, set as $\eta \propto \sqrt{\sigma^2 N_{\text{B}} \log(N_{\text{B}})}$~\cite{ZHANG201995}. It can be further expressed as 
\begin{align}\label{ANM_Theta_first}
\{ \hat{\tilde{\bu}}_1, \hat{\tilde{\bZ}}, \hat{\tilde{\bU}}\}=&\arg\min_{\tilde{\bu}_1,\tilde{\bZ},\tilde{\bU}} \frac{\eta}{2 M_0}\mathrm{Tr}(\tilde{\bZ}) + \frac{\eta}{2 N_{\text{B}}} \mathrm{Tr}(\mathrm{Toep}(\tilde{\bu}_1))\nonumber\\
& + \frac{1}{2}\|\bY_0^{\mathsf{H}}-  \bX_0^{\mathsf{H}}\tilde{\bU}\|_{\mathrm{F}}^2 \nonumber\\
&\text{s.t.} \;\begin{bmatrix} \mathrm{Toep}(\tilde{\bu}_1)  & \tilde{\bU}\\
\tilde{\bU}^{\mathsf{H}}& \tilde{\bZ}
\end{bmatrix} \succeq \mathbf{0}.
\end{align}
Similarly, the recovery of $\boldsymbol{\theta}_{\text{B,R}}$ is based on the solution of $\mathrm{Toep}(\hat{\tilde{\bu}}_1)$ from~\eqref{ANM_Theta_first} by root finding approach or other related approaches. 


\subsection{Second Stage of Channel Estimation Algorithm} \label{subsect:2nd_state}

In the second stage, we first design training and receive beams, which leads to a simplified approximate observation model. From this model, we can determine $L_{\text{B,R}} L_{\text{R,M}}$ separate observations and apply SMV atomic norm minimization on each of these. These different steps are now detailed. 

\subsubsection{Training and Receive Beams}

After estimation of $\boldsymbol{\theta}_\text{B,R}$ and $\boldsymbol{\phi}_\text{R,M}$, we align the training beams at BS and receiving beams at MS with these angles. Namely, we design the $\bX_t$ and $\bW_t$, for $t = 1,\cdots, T$, as follows 
\begin{align}
\bX_t &= \frac{1}{\sqrt{N_{\text{B}}}} \bA(\hat{\boldsymbol{\theta}}_\text{B,R}), \nonumber \\
\bW_t &= \frac{1}{\sqrt{N_{\text{M}}}} \bA(\hat{\boldsymbol{\phi}}_\text{R,M}),
\end{align}
where $\hat{\boldsymbol{\theta}}_\text{B,R}$ and $\hat{\boldsymbol{\phi}}_\text{R,M}$ are the estimates of $\boldsymbol{\theta}_\text{B,R}$ and $\boldsymbol{\phi}_\text{R,M}$, respectively, from the first stage. The numbers of columns in $\bX_t$ and $\bW_t$ are $L_{\text{B,R}}$ and $L_{\text{R,M}}$, respectively. In general, these values are far less than the number of the training beams/sequences used in the first stage, i.e., $L_{\text{B,R}} \ll N_0$ and $L_{\text{R,M}} \ll M_0$. Therefore, the training overhead can be reduced tremendously by first determining $\boldsymbol{\theta}_\text{B,R}$ and $\boldsymbol{\phi}_\text{R,M}$ in the first stage and then guiding the design of $\bX_t$ and $\bW_t$, used in the second stage. 

\subsubsection{Simplified Observation Model}

Assuming we have a very accurate estimate in the first stage, i.e., $\hat{\boldsymbol{\theta}}_\text{B,R} \approx \boldsymbol{\theta}_\text{B,R}$ and $\hat{\boldsymbol{\phi}}_\text{R,M} \approx \boldsymbol{\phi}_\text{R,M} $, we have the following 
\begin{align}\label{Product_array_train}
\bA^{\mathsf{H}}(\boldsymbol{\theta}_\text{B,R}) \bX_t &\approx \sqrt{N_{\text{B}}} \bI, \nonumber\\
\bW_t^{\mathsf{H}} \bA(\boldsymbol{\phi}_\text{R,M})  &\approx \sqrt{N_{\text{M}}} \bI,
\end{align}
under the condition of sufficient separation of angles and a large number of antennas at both BS and MS. In practice, the estimation performance depends on the SNR level, number of training sequences used in the first stage, and the size of the combining matrix in the first stage. \tcb{Super resolution estimation can be achieved in the high SNR regime with reasonable training overhead, as can be seen in the numerical study in Section~\ref{section_performance_evaluation}.} In general, the estimation in the first stage loses the order information on entries in $\boldsymbol{\theta}_\text{B,R}$ and $\boldsymbol{\phi}_\text{R,M}$. Therefore, the products may not be scaled identity matrices as in~\eqref{Product_array_train} but scaled elementary matrices. This does not affect the parameter estimation in the second stage, as explained in the sequel. 

Let us assume that the relationship in~\eqref{Product_array_train} holds. Then, the received signals in the second stage can be further approximated as
\begin{align}\label{Rec_signal2}
\bY_t  &=   \bW_t^{\mathsf{H}} \bA(\boldsymbol{\phi}_\text{R,M}) \bG_t\bA^{\mathsf{H}}(\boldsymbol{\theta}_\text{B,R})\bX_t + \bW_t^{\mathsf{H}} \bZ_t \nonumber \\
&\approx \sqrt{N_{\text{B}}N_{\text{M}}} \bG_t + \bW_t^{\mathsf{H}} \bZ_t , \;\text{for}\; t =1, \cdots, T.
\end{align}

\subsubsection{Formulation of $L_{\text{B,R}} L_{\text{R,M}}$ observations}

Recalling that $\bG_t =  \mathrm{diag}(\boldsymbol{\rho}_{\text{R,M}}) \bA^{\mathsf{H}}(\boldsymbol{\theta}_\text{R,M}) \boldsymbol{\Omega}_t\bA(\boldsymbol{\phi}_\text{B,R})\\ \mathrm{diag}(\boldsymbol{\rho}_{\text{B,R}})$, the $(m, n)$th entry of $\bG_t$ is in the form of 
\begin{align}\label{G_t_element}
    [\bG_t]_{mn} &= [\boldsymbol{\rho}_{\text{R,M}}]_m \boldsymbol{\omega}_t^{\mathsf{T}} \boldsymbol {\alpha}( [\boldsymbol{\Delta}]_{mn} ) [\boldsymbol{\rho}_{\text{B,R}}]_n,\nonumber\\
    & \text{for}\;\; m = 1,\cdots, L_{\text{R,M}}, n = 1,\cdots, L_{\text{B,R}},
\end{align}
where $[\boldsymbol{\Delta}]_{mn} = \mathrm{asin}\big(\sin([\boldsymbol{\phi}_{\text{B,R}}]_n)-\sin([\boldsymbol{\theta}_{\text{R,M}}]_m)\big)$ is the angle difference matrix associated with the RIS and $\boldsymbol{\omega}_t \in \mathbb{C}^{N_{\text{R}} \times 1}$ is the vector composed of diagonal elements of $\boldsymbol{\Omega}_t$, i.e., $\boldsymbol{\Omega}_t = \mathrm{diag}(\boldsymbol{\omega}_t)$. By setting $\bg_t = \mathrm{vec}(\bG_t)$, the $i$th element of $\bg_t$ is of the form of 
\begin{equation}\label{g_vectorization}
[\bg_t]_i =  \rho_i  \boldsymbol{\omega}_t^{\mathsf{T}} \boldsymbol {\alpha}(\tilde{\theta}_i)\;\; \text{for}\;\; i =1,\cdots,L_{\text{B,R}} L_{\text{R,M}},
\end{equation}
where 
\begin{align}\label{relationship_rho_theta}
    &\rho_i = [\boldsymbol{\rho}_{\text{R,M}}]_m [\boldsymbol{\rho}_{\text{B,R}}]_n, \nonumber\\
     &\tilde{\theta}_i = \mathrm{asin}\big( \sin([\boldsymbol{\phi}_{\text{B,R}}]_n)-\sin([\boldsymbol{\theta}_{\text{R,M}}]_m)\big), \nonumber\\
      &\text{with}\; m = (i-1)\% L_{\text{R,M}} +1, n = \Big\lceil \frac{i}{L_{\text{R,M}}}\Big\rceil,
\end{align}
where $\%$ is the modulo operation. In other words, the product of propagation path gains $\rho_i$ is taken from entries of vector $\boldsymbol{\rho} = \boldsymbol{\rho}_{\text{R,M}}\otimes \boldsymbol{\rho}_{\text{B,R}}$, and $\tilde{\theta}_i$ is taken from the set of angle differences
\begin{align}\label{Ang_diff}
\tilde{\Theta} =\{ \tilde{\theta}:  \mathrm{asin}\big( \sin( [\boldsymbol{\phi}_{\text{B,R}}]_n)-\sin([\boldsymbol{\theta}_{\text{R,M}}]_m)\big) ,\nonumber\\
 \; m = 1,\cdots, L_{\text{R,M}}, \; n =1,\cdots,  L_{\text{B,R}}\}.
\end{align}
Therefore, each element in $\mathrm{vec}(\bY_t)$ corresponds to one couple of unknown parameters $\{\rho_i,\tilde{\theta}_i\}$, $i=1,\ldots, L_{\text{B,R}} L_{\text{R,M}}$. We now gather these observations across $T$ transmission blocks. By introducing $\bY = \big[\mathrm{vec}(\bY_1), \cdots, \mathrm{vec}(\bY_T)\big]$ and $\bar{\bG} = [\bg_1, \cdots, \bg_T]$, each element in the $i$th row in $\bY$, denoted by $[\bY]_{i,:}$, corresponds to the same $\{\rho_i,\tilde{\theta}_i\}$. Hence, we can express the $i$th row in column format as
\begin{align}\label{signal_at_second}
[\bY]_{i,:}^{\mathsf{T}} & \approx \sqrt{N_{\text{B}}N_{\text{M}}} [\bar{\bG}]_{i,:}^{\mathsf{T}} +  \bz_i, \nonumber\\
&=  \sqrt{N_{\text{B}}N_{\text{M}}}  [\boldsymbol{\omega}_1, \cdots, \boldsymbol{\omega}_T]^{\mathsf{T}} \rho_i\boldsymbol {\alpha}(\tilde{\theta}_i) + \bz_i, \nonumber\\
&= \sqrt{N_{\text{B}}N_{\text{M}}}  \bar{\boldsymbol{\Omega}} \rho_i \boldsymbol{\alpha}(\tilde{\theta}_i) + \bz_i,
\end{align}
where $\bar{\boldsymbol{\Omega}} = [\boldsymbol{\omega}_1, \cdots, \boldsymbol{\omega}_T]^{\mathsf{T}}$ and $\bz_i$ is the additive noise as $\bz_i = [\mathrm{vec}(\bW_1^{\mathsf{H}} \bZ_1), \cdots, \mathrm{vec}(\bW_T^{\mathsf{H}} \bZ_T)]_{i,:}^{\mathsf{T}}$. 

\subsubsection{SMV Atomic Norm Minimization}
According to the formulation~\eqref{signal_at_second}, this incurs $L_{\text{B,R}} L_{\text{R,M}}$ sparsity-1 signal recovery problems with $\bar{\boldsymbol{\Omega}}$ being the linear measurement matrix. We can estimate $\rho_i$ and $\tilde{\theta}_i$ by resorting to atomic norm minimization on SMV. It should be noted that we cannot estimate $\boldsymbol{\rho}_{\text{R,M}}$ and $\boldsymbol{\rho}_{\text{B,R}}$ separately due to the coupling effect, and the same principle applies to $\boldsymbol{\phi}_{\text{B,R}}$ and $\boldsymbol{\theta}_{\text{R,M}}$, as seen in~\eqref{G_t_element} and~\eqref{relationship_rho_theta}.

In the second stage, $L_{\text{B,R}} L_{\text{R,M}}$ atomic norm minimization problems are formulated as 
\begin{align}\label{ANM_Second_stage}
\{ \hat{\bv}, \hat{\bh}_i, \hat{z}\}=&\arg\min_{\bv,\bh_i,z} 0.5\nu_i z 
+ \frac{\nu_i}{2 N_{\text{R}}} \mathrm{Tr}(\mathrm{Toep}(\bv)) \nonumber\\
&+ \frac{1}{2}\|[\bY]_{i,:}^{\mathsf{T}} -\sqrt{N_{\text{B}}N_{\text{M}}}\bar{\boldsymbol{\Omega}}\bh_i\|_2^2 \nonumber\\
&\text{s.t.} \;\begin{bmatrix} \mathrm{Toep}(\bv)  & \bh_i\\
\bh_i^{\mathsf{H}}& z
\end{bmatrix} \succeq \mathbf{0}, \; \text{for}\; i = 1,\cdots, L_{\text{B,R}} L_{\text{R,M}},
\end{align}
where $\bh_i = \rho_i\boldsymbol {\alpha}(\tilde{\theta}_i)$ and the regularization parameter $\nu_i$ is set as $\nu_i \propto \sqrt{\sigma^2 N_{\text{R}} \log(N_{\text{R}})}$. The estimate of $\tilde{\theta}_i$, denoted as $\hat{\tilde{\theta}}_i$, relies on $\mathrm{Toep}(\hat{\bv})$ by resorting to root finding methods. The estimation of $\rho_i$ is obtained by using least squares (LS) as 
\begin{equation}
    \hat{\rho}_i = \big(\boldsymbol {\alpha}(\hat{\tilde{\theta}}_i)\big)^{\dagger} \hat{\bh}_i,
\end{equation}
where $(\cdot)^{\dagger}$ denotes Moore-Penrose pseudo-inverse and $\hat{\bh}_i$ is the solution from~\eqref{ANM_Second_stage} for $\bh_i$. 

\begin{figure}[t]
	\centering
	\includegraphics[width=0.5\linewidth]{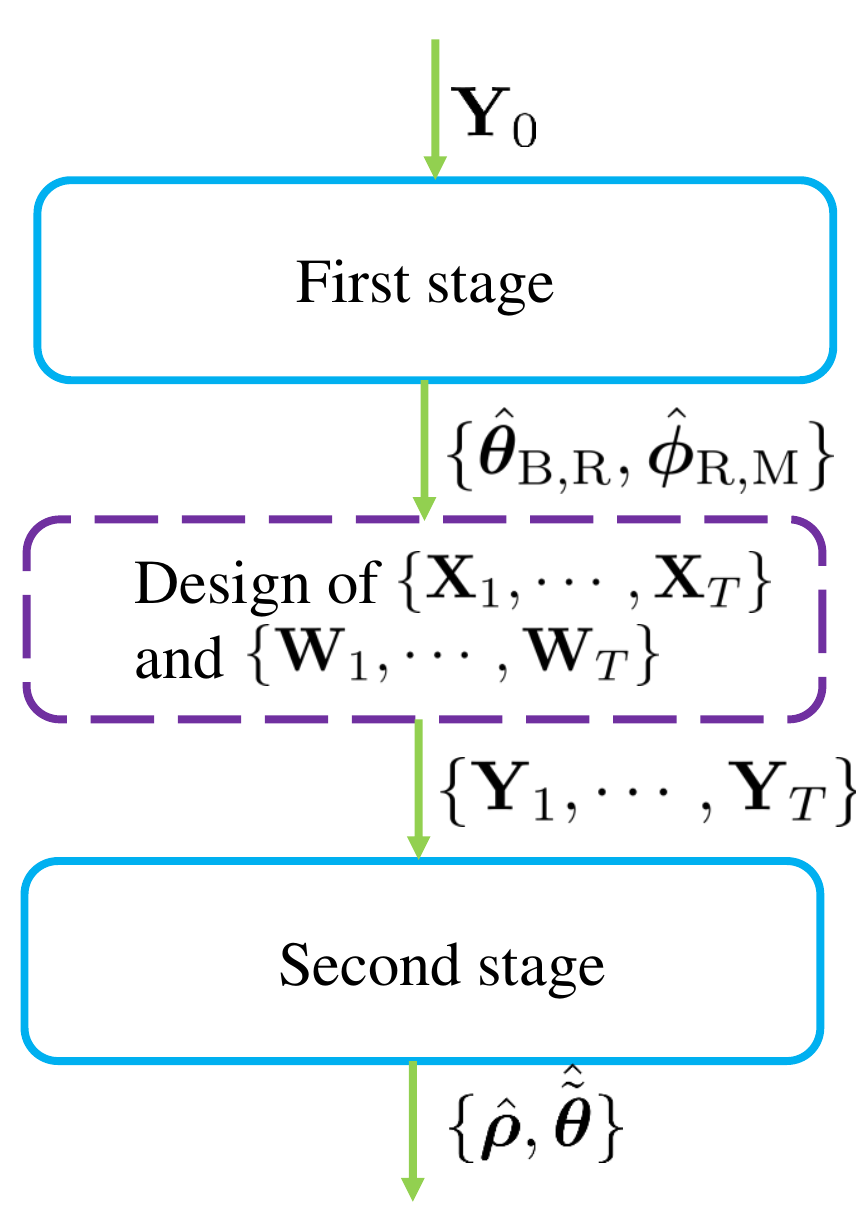}
	\caption{The proposed two-stage CE approach, where in the first stage AoDs from BS-RIS channel and AoAs from RIS-MS channel are determined and in the second stage, training and receive beams aligned with these directions are used to collect observations to estimate the products of propagation path gains and angle differences.  }
	\label{Two_stage_CE}
\end{figure}

The proposed two-stage CE approach is summarized in Fig.~\ref{Two_stage_CE}.

\begin{remark}
There exists one-to-one correspondence between $\{\rho_i, \tilde{\theta}_i\}$ and $[\bY]_{i,:}$, depicted in~\eqref{signal_at_second}. As shown in~\eqref{ANM_Second_stage}, we estimate the parameter pairs $\{\rho_i, \tilde{\theta}_i\}$ one by one based on one row from $\bY$. The loss of order information on entries in $\boldsymbol{\theta}_\text{B,R}$ and $\boldsymbol{\phi}_\text{R,M}$ in the first CE stage will only change the row order of $\bY$ accordingly, which will only changes the order of estimating the parameter pairs other than bring negative effect on the estimation accuracy. 
\end{remark}

%
%
\subsection{Complexity Analysis and Training Overhead}\label{CA_TO}
The computational complexity in the first stage depends on the size of the positive semidefinite matrix in~\eqref{ANM_Phi_first} and~\eqref{ANM_Theta_first}, i.e., $\max\Big\{O\big((N_{\text{B}}+M_0)^{3.5}\big), O\big((N_{\text{M}}+N_0)^{3.5}\big)\Big\}$~\cite{ZHANG201995}. In the second stage, the computational complexity is proportional to $O\big((N_{\text{R}}+1)^{3.5}\big)$. Therefore, the overall complexity is proportional to $\max\Big\{O\big((N_{\text{B}}+M_0)^{3.5}\big), O\big((N_{\text{M}}+N_0)^{3.5}\big), O\big((N_{\text{R}}+1)^{3.5}\big)\Big\}$, which is determined by the largest number among the three-tuple $\{ N_{\text{B}}+M_0, N_{\text{M}}+N_0, N_{\text{R}}+1\}$.
The overall training overhead is 
\begin{align}
    T_t = N_0 \Big\lceil \frac{M_0}{N_{\text{RF}}}\Big\rceil + T L_{\text{B,R}} \Big\lceil \frac{L_{\text{R,M}}}{N_{\text{RF}}}\Big\rceil. \label{eq:trainingoverhead}
\end{align}


\section{RIS Control and Beamforming \& Combining Design} \label{sec:Control}
The ultimate motivation of estimating the channel parameters discussed above is to enable coherent demodulation, to be able to design the phase control matrix at the RIS and transmit and receive beamforming vectors in order to maximize the SE.

\subsection{Design of $\boldsymbol{\Omega}$}\label{seciton_design_RIS}
 The optimization criterion used here is to maximize the power of $\bG$, defined in~\eqref{G_matrix}, as a function of $\boldsymbol{\Omega}$, i.e., $\|\bG\|_{\mathrm{F}}^2$, to maximize the effective SNR at the receiver.\footnote{\tcb{Note that the design of phase control matrix in~\eqref{Design_G} is heuristic, and not guaranteed to be optimal for SE maximization. Better criteria may exist for the design of $\boldsymbol{\Omega}$, which is left for our future investigation.}} The optimal design of $\boldsymbol{\Omega}$ is expressed as 
\begin{equation}\label{Design_G}
{\boldsymbol{\Omega}}^{\star} = \arg\max_{\boldsymbol{\Omega}} \|\bG\|_{\mathrm{F}}^2,
\end{equation}
where $ \|\bG\|_{\mathrm{F}}^2$ can be expressed as 
\begin{align}
 &\|\bG\|_{\mathrm{F}}^2 = \| \mathrm{diag}(\boldsymbol{\rho}_{\text{R,M}}) \bA^{\mathsf{H}}(\boldsymbol{\theta}_\text{R,M}) \boldsymbol{\Omega}\bA(\boldsymbol{\phi}_\text{B,R}) \mathrm{diag}(\boldsymbol{\rho}_{\text{B,R}})\|_{\mathrm{F}}^2 \nonumber\\
& \overset{(a)}{=}  \sum_{ n = 1}^{L_{\text{B,R}} } \sum_{ m= 1}^{L_{\text{R,M}}} \Big| [\boldsymbol{\rho}_{\text{B,R}}]_n [\boldsymbol{\rho}_{\text{R,M}}]_m \boldsymbol{\omega}^{\mathsf{T}} \big(\boldsymbol {\alpha}^*([\boldsymbol{\theta}_{\text{R,M}}]_m) \circ \boldsymbol {\alpha}([\boldsymbol{\phi}_{\text{B,R}}]_n)\big) \Big|^2 \nonumber\\
& \overset{(b)}{=}\sum_{ i = 1}^{L_{\text{B,R}} L_{\text{R,M}}} \Big|  \rho_i\boldsymbol{\omega}^{\mathsf{T}} \boldsymbol {\alpha}(\tilde{\theta}_i)   \Big|^2,
\end{align}
where $(a)$ and $(b)$ are obtained by following~\eqref{G_t_element} and~\eqref{g_vectorization}, respectively, and $\boldsymbol{\omega} = \mathrm{diag}(\boldsymbol{\Omega})$. Therefore, the optimal $\boldsymbol{\omega}$ (denoted by $\boldsymbol{\omega}^{\star}$) based on the estimates in the second stage is obtained by 
\begin{align}
{\boldsymbol{\omega}^{\star}} &= \arg\max_{\boldsymbol{\omega}} \sum_{ i = 1}^{L_{\text{B,R}} L_{\text{R,M}}}  \Big|  \hat{\rho}_i\boldsymbol{\omega}^{\mathsf{T}} \boldsymbol {\alpha}(\hat{\tilde{\theta}}_i)  \Big |^2\nonumber\\
& = \arg\max_{\boldsymbol{\omega}} \boldsymbol{\omega}^{\mathsf{T}} \bE \bE^{\mathsf{H}} \boldsymbol{\omega}^*,
\end{align}
where 
\begin{align}
    \bE = [\boldsymbol {\alpha}(\hat{\tilde{\theta}}_1), \cdots, \boldsymbol {\alpha}(\hat{\tilde{\theta}}_{L_{\text{B,R}} L_{\text{R,M}}})] \mathrm{diag}([\hat{\rho}_1, \cdots, \hat{\rho}_{L_{\text{B,R}} L_{\text{R,M}}}] ).
\end{align}
We conduct singular value decomposition (SVD) on $\bE \bE^{\mathsf{H}}$ as $\bE \bE^{\mathsf{H}} = \bJ\bD\bJ^{\mathsf{H}}$, where $\bJ\bJ^{\mathsf{H}} = \bJ^{\mathsf{H}} \bJ = \bI$ and $\bD$ is a diagonal matrix with singular values on the diagonal as a descending order. The optimal $\boldsymbol{\omega}^{\star}$ is chosen as the conjugate of the first column of $\bJ$ and then projected to the unit-modulus vector space, i.e., $ \boldsymbol{\omega}^{\star} = \exp(-j\mathrm{phase}([\bJ]_{:,1}))$, where $\mathrm{phase}(\cdot)$ denotes the element-wise operation of extracting the phases of the argument.  

\begin{remark}
The optimal phase control matrix ${\boldsymbol{\Omega}^{\star}} = \mathrm{diag}({\boldsymbol{\omega}^{\star}})$ for the power maximization criterion of the effective channel is closely aligned with the conjugate of the singular vector associated with the largest singular value of the matrix $\bE \bE^{\mathsf{H}}$. 
\end{remark}

\subsection{Beamforming at BS and Combining at MS}\label{BF_at_BS_MS}
The BS BF and MS combining design is based on the estimate of composite channel after setting $\boldsymbol{\Omega}^{\star} = \mathrm{diag}(\boldsymbol{\omega}^{\star})$. The reconstructed composite channel is formulated as 
\begin{equation}\label{H_based_on_estimates}
\hat{\bH} = \bA(\hat{\boldsymbol{\phi}}_\text{R,M}) \hat{\bG} \bA^{\mathsf{H}}(\hat{\boldsymbol{\theta}}_\text{B,R}),
\end{equation}
where $\hat{\bG} = \mathrm{vec2mat}(\hat{\bg})$ with $ [\hat{\bg}]_i = \hat{\rho}_i\boldsymbol{\omega^{\star}}^{\mathsf{T}} \boldsymbol {\alpha}(\hat{\tilde{\theta}}_i)$, constructed by using $\boldsymbol{\Omega}^{\star}$ and estimates in the second stage, i.e., $\{\hat{\rho}_i, \hat{\tilde{\theta}}_i\}$, and $\mathrm{vec2mat}(\cdot)$ converts a vector to a matrix with a predefined size.\footnote{Here, $\mathrm{vec2mat}(\cdot)$ is an inverse operation of $\mathrm{vec}(\cdot)$. For instance, we have $\hat{\bg} = \mathrm{vec}(\hat{\bG})$, and on the contrary, we have $\hat{\bG} = \mathrm{vec2mat}(\hat{\bg})$ under the condition that the size of $\hat{\bG}$ is known.} The SVD is further applied to $\hat{\bH}$ as $\hat{\bH} = \breve{\bU} \boldsymbol{\Sigma} \breve{\bV}^{\mathsf{H}}$, and the optimal BF and combining vectors at the BS and MS are aligned with the singular vectors associated with the largest singular value, i.e., the BF vector at the BS as $\bf \approx [\breve{\bV}]_{:,1}$ and the combining vector at the MS as $\bw \approx [\breve{\bU}]_{:,1}$ after taking into consideration the constraints of the hybrid precoding architecture.\footnote{We use $\approx$ here due to the inherent hardware constraints, which may bring some gap between $\bf (\bw)$ and $[\breve{\bV}]_{:,1} ([\breve{\bU}]_{:,1})$. If no constraints exist, like that in the full digital precoding systems, $=$ will be used instead.}

\section{Performance Evaluation}\label{section_performance_evaluation}
In this section, we demonstrate the efficiency of the proposed CE approach. We present several benchmarks, detail the simulation scenario parameters as well as performance metrics, and provide an in-depth performance analysis and discussion. 
\subsection{Benchmarks}
For the benchmark scheme, we consider the OMP based two-stage approach. In the first stage, the vectorization of $\bY_0$ is in the form of
\begin{align} \label{vec_rec_signal}
\by_0 &= \mathrm{vec}(\bY_0) = (\bX_0^{\mathsf{T}} \otimes \bW_0^{\mathsf{H}}) \mathrm{vec}(\bH(\boldsymbol{\Omega}_0)) +  \mathrm{vec}(\bW_0^{\mathsf{H}} \bZ_0),\nonumber\\
& =  (\bX_0^{\mathsf{T}} \otimes \bW_0^{\mathsf{H}})\bar{\bA} \bg_0 +  \bn_0,
\end{align}
where $\bar{\bA} = \bA^{*}(\boldsymbol{\theta}_\text{B,R}) \otimes\bA(\boldsymbol{\phi}_\text{R,M}) $ and
$\bn_0 = \mathrm{vec}(\bW_0^{\mathsf{H}} \bZ_0)$. $\bar{\bA} \bg_0$ in \eqref{vec_rec_signal} can be further expressed as $\bar{\bA} \bg_0 = \bA_\text{d} \tilde{\bg}_0$, where $\bA_\text{d}$ is deemed as an overcomplete dictionary containing the columns of $\bar{\bA}$ and constructed by quantizing the angular domains of AoD of the BS-RIS channel and AoA of RIS-MS channel into $2 N_{\text{B}}$ and $2 N_{\text{M}}$ levels, respectively. Ideally, $\tilde{\bg}_0$ is a vector with $L_{\text{B,R}}L_{\text{R,M}}$ elements the same as these of $\bg_0$ while the remaining elements are all-zeros. In other words, $\bar{\bA} \bg_0$ can be sparsely represented under a certain overcomplete dictionary. $\bX_0^{\mathsf{T}} \otimes \bW_0^{\mathsf{H}}$ is considered as the linear measurement matrix. \tcb{Therefore, the recovery of $\bar{\bA}$ (or equivalently $\boldsymbol{\theta}_\text{B,R}$ and $\boldsymbol{\phi}_\text{R,M}$) and $\bg_0$ can be addressed by resorting to the OMP algorithm~\cite{Tropp2007}, which sequentially finds the atoms from the overcomplete dictionary $\bA_\text{d}$ in order to greedily improve the approximation.} In the second stage, the dictionary is constructed by quantizing the angular domains into $2 N_{\text{R}}$ and each atom is in the form of an array response vector. The recovery of $\{\rho_i, \tilde{\theta}_i\}$ is also conducted by using OMP on~\eqref{signal_at_second}.

We also consider two benchmarks under perfect CSI:
(i) CSI of the individual channels is perfectly known to evaluate the average SE, \tcb{where the RIS phase control matrix, BS beamformer, and MS combiner are jointly designed via an iterative method}. This perfect CSI may be obtained by knowing the exact location information of the BS, MS, and RIS and environmental information~\cite{Taranto2014};
    (ii) CSI of the LoS path is perfectly known, where we align the beams with the angles related to the LoS path and evaluate the average SE bound.

\subsection{System Parameters and Performance Metrics} \label{subsec:sys_para}
The simulation parameters are set as follows: $N_{\text{B}} = N_{\text{M}} = 16$, \tcb{$N_{\text{R}} = 64$}, and $N_{\text{RF}} = 8$. The angle separation in terms of directional sine is assumed to be larger than ${4}/{N_{\text{B}}}$, ${4}/{N_{\text{R}}}$, and ${4}/{N_{\text{M}}}$ at the BS, RIS, and MS, respectively.   
We assume that the propagation path gains follow $\mathcal{CN}(0,1)$ until Section~\ref{Section_effect_path_gain} and each element of $\bZ_t$ follows $\mathcal{CN}(0,\sigma^2)$. The SNR is defined as ${1}/{\sigma^2}$, and 2000 realizations are considered for averaging. \tcb{Without loss of generality, we fix the channel coherence time as $500$ (in symbol times, i.e., $T_c = 500$) in the evaluation of effective SE bound.} 

Performance will be assessed in several metrics: (i) the MSE of the estimated parameters (angles in the first stage,  angle difference and the product of propagation path gains in the second stage), (ii) the average effective SE bound; (iii)
the average squared distance (ASD) between the designed beamformer (combiner) in Section~\ref{BF_at_BS_MS} and the optimal one obtained by assuming full CSI; and (iv)
 the RIS gain based on the estimated parameters.  
The MSEs of angular parameter estimation and product of propagation path gains estimation are defined as\footnote{Another way to formulate the MSEs is directly based on the angular estimates without taking sine operation. Nevertheless, the results based on the two types of calculations will be consistent.}
\begin{align}
    \mathrm{MSE}\big(\sin(\boldsymbol{\theta}_{\text{B,R}})\big) &= \mathbb{E}\Big[\frac{\|\sin(\boldsymbol{\theta}_{\text{B,R}}) - \sin(\hat{\boldsymbol{\theta}}_{\text{B,R}})\|^2_2}{L_{\text{B,R}}}\Big],\nonumber\\
     \mathrm{MSE}\big(\sin(\boldsymbol{\phi}_{\text{R,M}})\big) &= \mathbb{E}\Big[\frac{\|\sin(\boldsymbol{\phi}_{\text{R,M}}) - \sin(\hat{\boldsymbol{\phi}}_{\text{R,M}})\|^2_2}{L_{\text{R,M}}}\Big],\nonumber\\
      \mathrm{MSE}\big(\sin(\boldsymbol{\Delta})\big) &= \mathbb{E}\Big[\frac{\|\sin(\boldsymbol{\Delta}) - \sin(\hat{\boldsymbol{\Delta}})\|^2_{\mathrm{F}}}{L_{\text{B,R}}L_{\text{R,M}}}\Big],\nonumber\\
         \mathrm{MSE}(\boldsymbol{\rho}) &= \mathbb{E}\Big[\frac{\|\boldsymbol{\rho} - \hat{ \boldsymbol{\rho}} \|^2_2}{L_{\text{B,R}}L_{\text{R,M}}}\Big].
\end{align}

The average effective SE bound for a given channel realization is defined as\footnote{It should be noticed that this is an asymptotic theoretical lower bound on data rate for the subsequent data transmission phase after designing the beamformers $\bw$ and $\bf$ and RIS phase control matrix $\boldsymbol{\Omega}^{\star}$, based on the estimates by the proposed CE scheme.}~\cite{Ngo2013}
\begin{equation}\label{effective_SE_Eq}
    R \hspace{-0.05cm}=\hspace{-0.05cm}  \mathbb{E} \hspace{-0.05cm}\Bigg[ \frac{T_c-T_t}{T_c}\log_2\hspace{-0.1cm}\bigg(1 +  \frac{|\bw^{\mathsf{H}} \hat{\bH} \bf|^2}{\sigma^2 \hspace{-0.05cm}+\hspace{-0.05cm}\mathrm{var}\big(\bw^{\mathsf{H}} \bH_{\text{e}}(\boldsymbol{{\Omega}^{\star}}) \bf\big) }\hspace{-0.1cm}\bigg)\hspace{-0.1cm}\Bigg] \, \text{bits/s/Hz}, 
\end{equation}
where the design of $\boldsymbol{{\Omega}^{\star}}$ was discussed in Section~\ref{seciton_design_RIS} and the design of $\bw$ and $\bf$ in Section~\ref{BF_at_BS_MS}, and  $\bH_{\text{e}}(\boldsymbol{{\Omega}^{\star}})$ is the channel estimation error, defined as $\bH_{\text{e}}(\boldsymbol{{\Omega}^{\star}}) = \bH(\boldsymbol{{\Omega}^{\star}}) - \hat{\bH}$. Recall that $T_c$ denotes that number of time slots in a coherence time interval, while $T_t$ is the training time from \eqref{eq:trainingoverhead}, expressed as a multiple of the OFDM symbol duration. \tcb{As can be seen in~\eqref{effective_SE_Eq}, the average effective SE bound is closely coupled with the estimation accuracy, the training overhead, and the design criterion of joint active and passive beamformers. Therefore, we also introduce it here as a performance metric, like in~\cite{Hu2018TVT, Alkhateeb2014,Taha2019}.} As said above, we average the SE results over 2000 channel realizations.  

The ASD of the beamformer is defined as
\begin{align}
    \mathrm{ASD}_f & = \mathbb{E}[\Vert \mathbf{f}-\mathbf{f}_o\Vert^2_2],\\
    \mathrm{ASD}_w & = \mathbb{E}[\Vert \mathbf{w}-\mathbf{w}_o\Vert^2_2],
\end{align}
 where $\bf_o$ and $\bw_o$ denote the optimal beamformer and combiner at the BS and MS, respectively (assuming full CSI).

Finally,
the RIS gain is defined as 
\begin{equation}
G_{\text{RIS}} = | \bA^{\mathsf{H}}(\boldsymbol{\theta}_{\text{R,M}}) \boldsymbol{{\Omega}^{\star}} \bA(\boldsymbol{\phi}_{\text{B,R}}) |_{\mathrm{F}}^2/N^2_{\text{R}}.
\end{equation}

\subsection{Results and Discussion}
\subsubsection{Single Path Scenario}
\begin{figure}[t]
	\centering
\includegraphics[width=.55\linewidth,trim={0 0 0 3},clip]{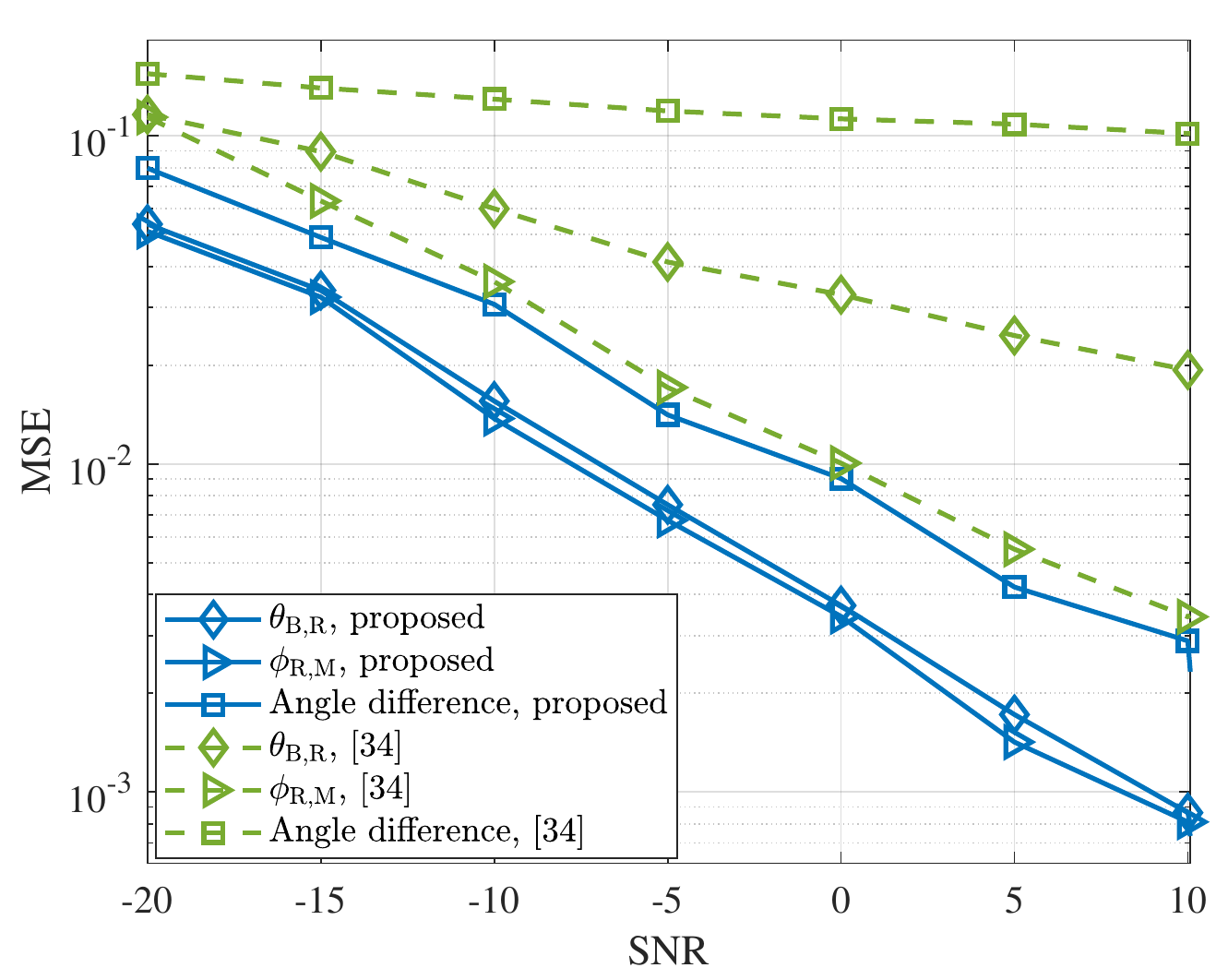}
	\caption{\tcb{Comparison between the proposed CE algorithm and that from~\cite{he2020channel} in terms of channel parameter estimation for the single path scenario.}}
	\label{parameter_estimate_comparison}
\end{figure}

\tcb{As an initial study, we make a comparison between our proposed CE scheme with that from~\cite{he2020channel} (using iterative reweighted method) for single path scenario, i.e., $L_{\text{B,R}} = L_{\text{R,M}} =1$. The simulation results are provided in Fig.~\ref{parameter_estimate_comparison}, where the training overhead is $T_t = 30$. As seen from the figure, the proposed CE scheme outperforms that in~\cite{he2020channel}.}
\subsubsection{Effect of Training Overhead}
\begin{figure}[t]
	\centering
\includegraphics[width=.55\linewidth,trim={0 0 0 3},clip]{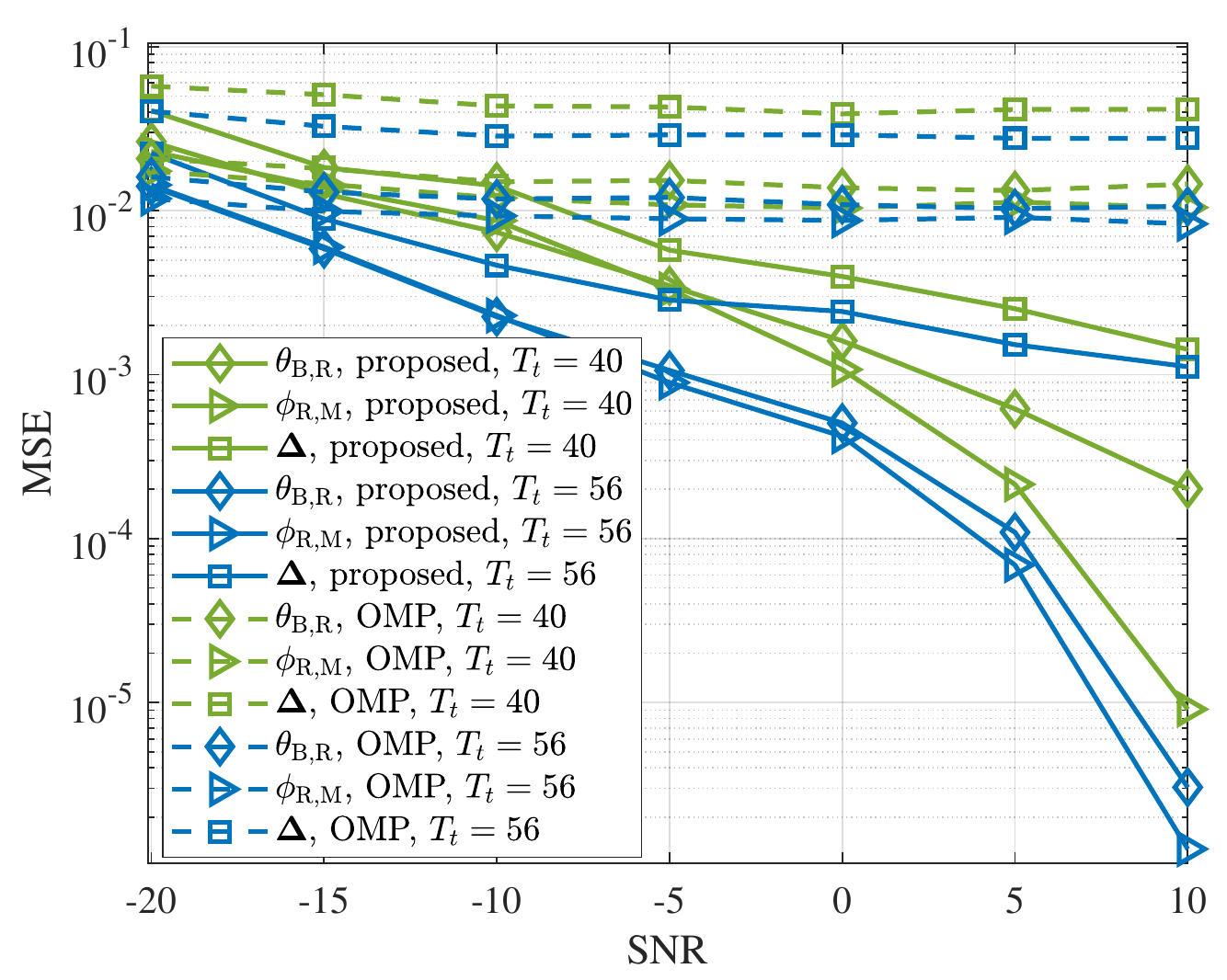}
	\caption{\tcb{The effect of training overhead on angular parameter estimation performance.}}
	\label{Channel_parameter_estimation}
\end{figure}

\begin{figure}[t]
	\centering
\includegraphics[width=.55\linewidth,trim={0 0 0 3},clip]{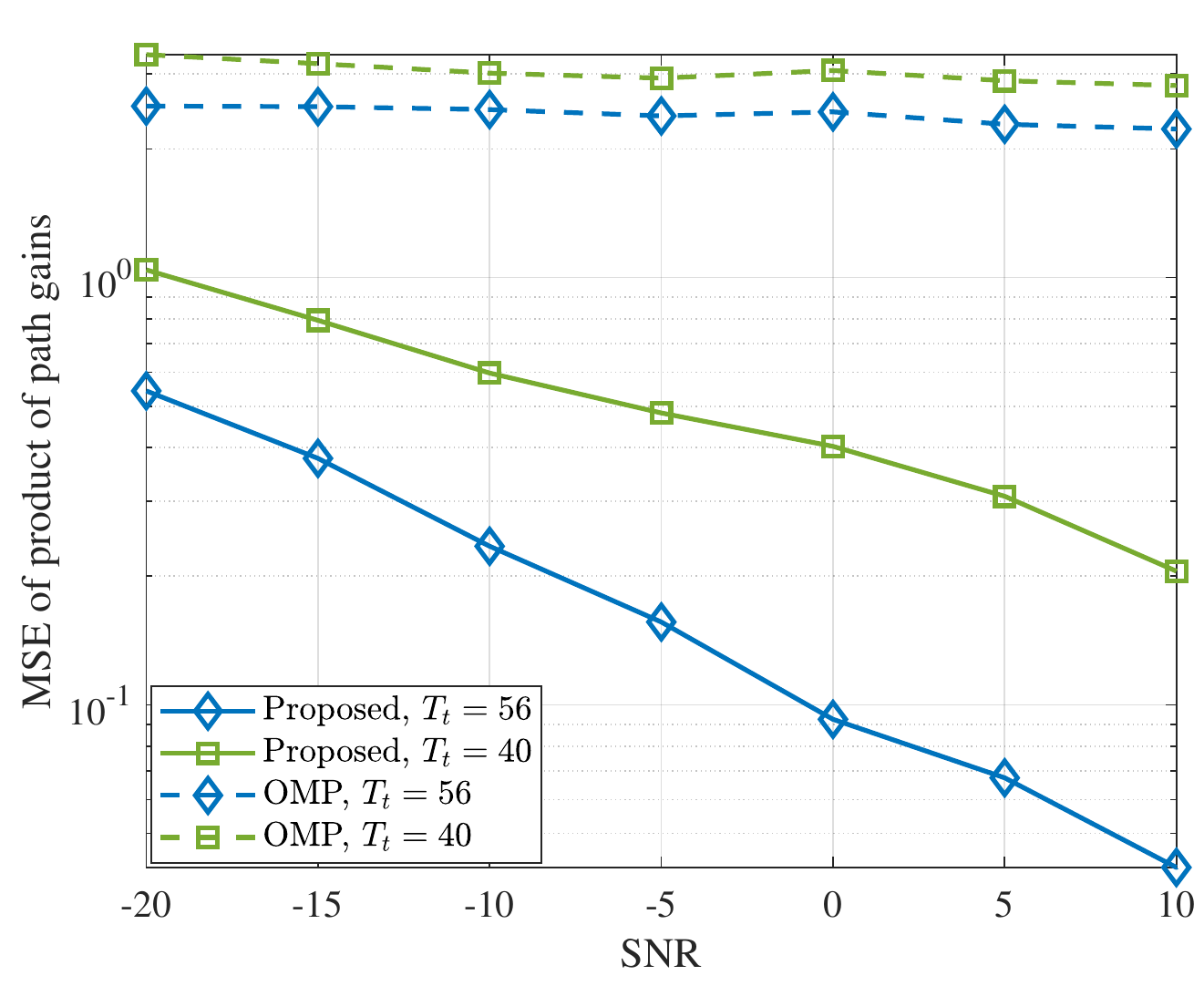}
	\caption{\tcb{The effect of training overhead on product of propagation path gains estimation performance in the second stage.}}
	\label{path_gain_estimation}
\end{figure}

\begin{figure}[t]
	\centering
	\includegraphics[width=.55\linewidth,trim={0 0 3 6},clip]{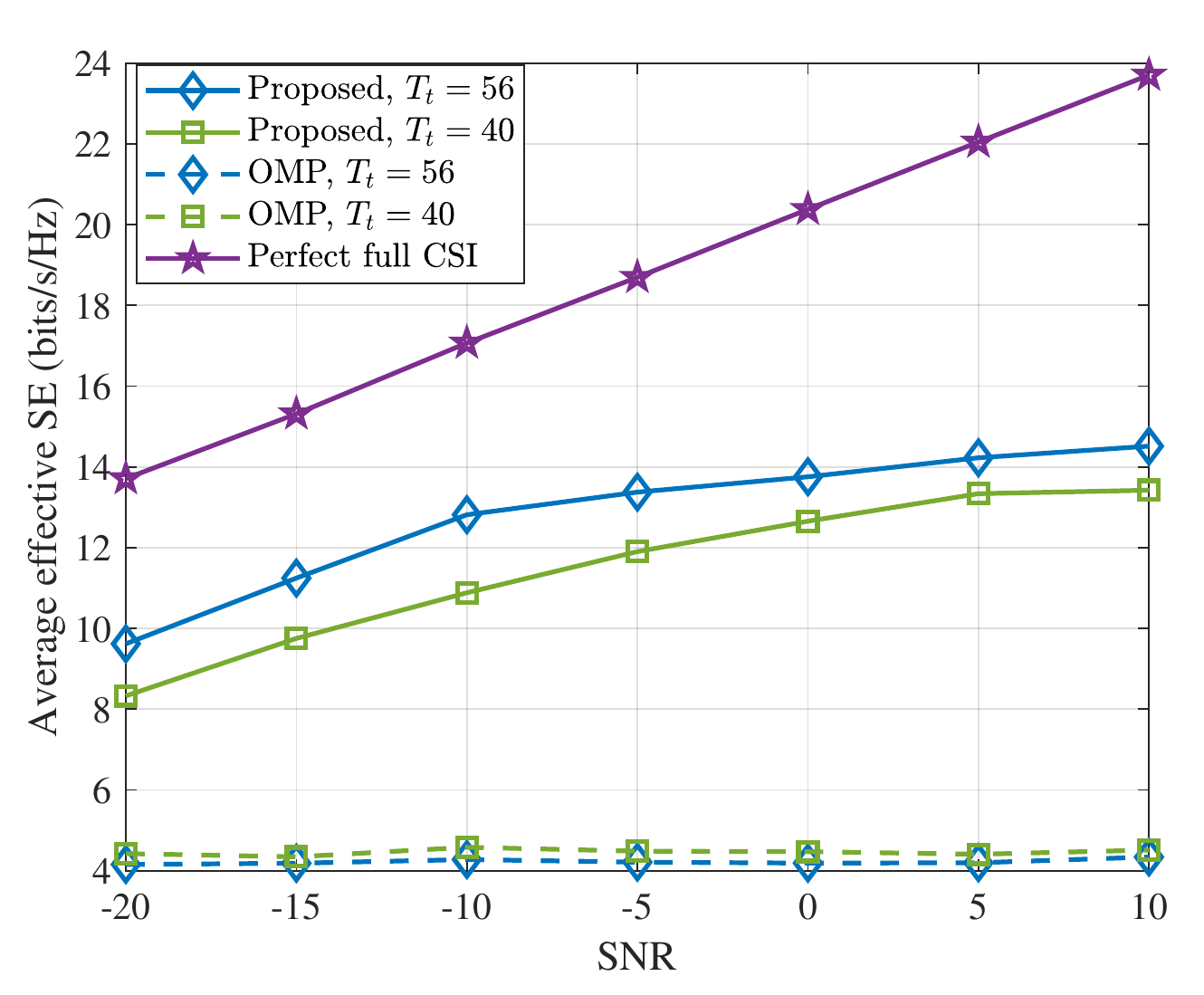}
	\caption{\tcb{Average effective SE bound vs.\ SNR.}}
	\label{Capacity_based_on_estimates}
\end{figure}

\begin{figure}[t]
	\centering
	\includegraphics[width=.55\linewidth,trim={0 0 3 6},clip]{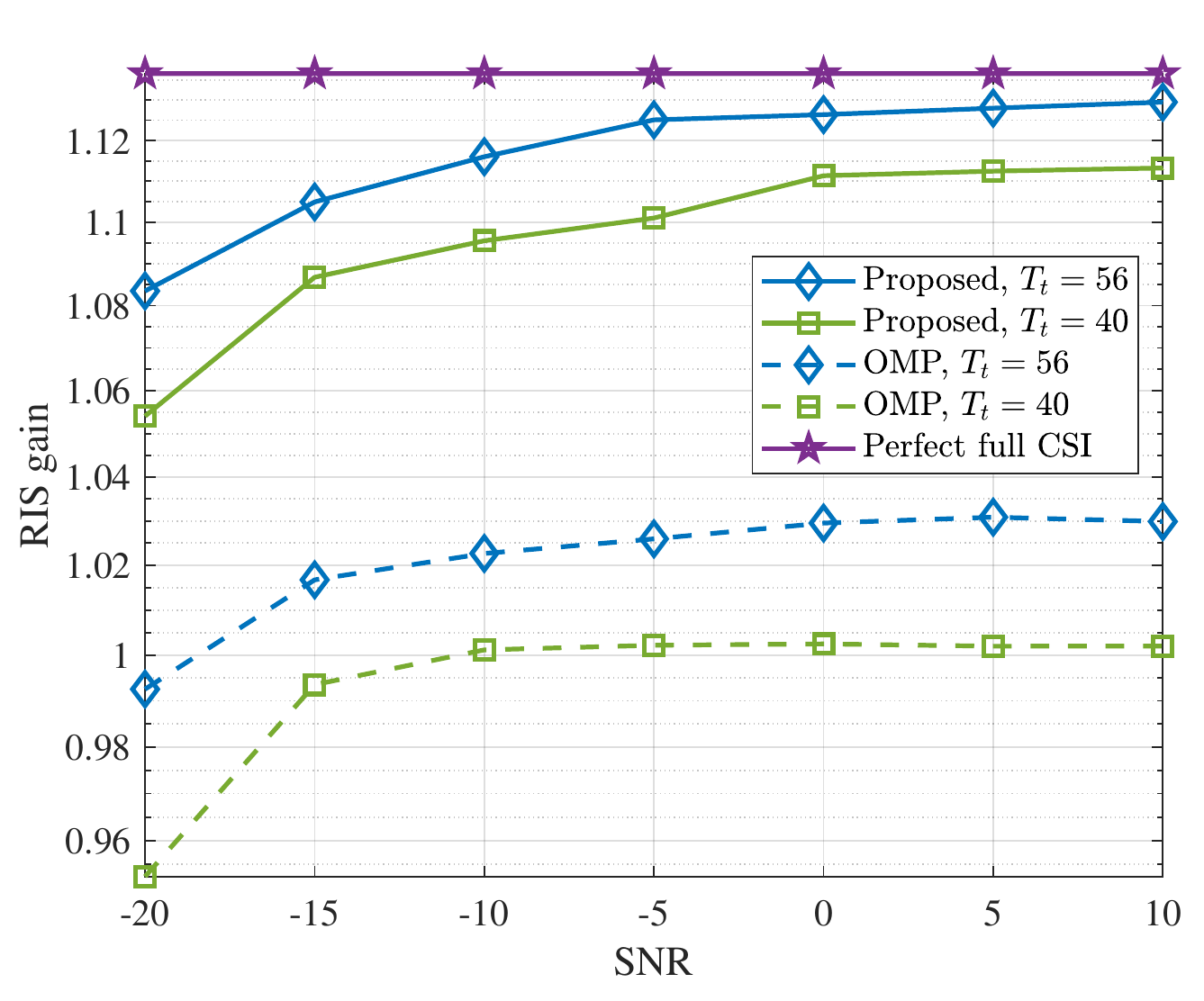}
	\caption{\tcb{RIS gain vs.\ SNR.}}
	\label{RIS_gain}
\end{figure}

The simulation results on the impact of training overhead on the parameter estimation performance as a function of the SNR are shown in Figs.~\ref{Channel_parameter_estimation} and~\ref{path_gain_estimation} for $L_{\text{B,R}} = L_{\text{R,M}} =2$ with two different setups: $N_0 = M_0 = T= 10$ with $T_t =40$ and  $N_0 = M_0 = T = 14$ with $T_t =  56$. The results in Figs.~\ref{Channel_parameter_estimation} and~\ref{path_gain_estimation} show that the increasing training overhead brings better performance on the channel parameter estimation at both stages as expected. The angular parameter estimation performance of the OMP-based benchmark scheme saturates to the level of $10^{-2}$ while the proposed scheme can bring better performance even in the low SNR regime, where a mild saturation of our scheme can also be observed. The results for the average effective SE bound and RIS gains are provided in Figs.~\ref{Capacity_based_on_estimates} and~\ref{RIS_gain}, which are aligned with the results for channel parameter estimates, shown in Figs.~\ref{Channel_parameter_estimation} and~\ref{path_gain_estimation}. The proposed scheme approximates the perfect CSI case in the low SNR regime with only dozens of consumed time slots in terms of both average effective SE bound and RIS gains. The saturation of average effective SE bound may results from the saturation of variance of channel estimation error, and this phenomenon needs to be further studied in depth as our future work. 

\subsubsection{Effect of Path Gain Profile}\label{Section_effect_path_gain}
\begin{figure}[t]
	\centering
	\includegraphics[width=.55\linewidth,trim={0 0 3 6},clip]{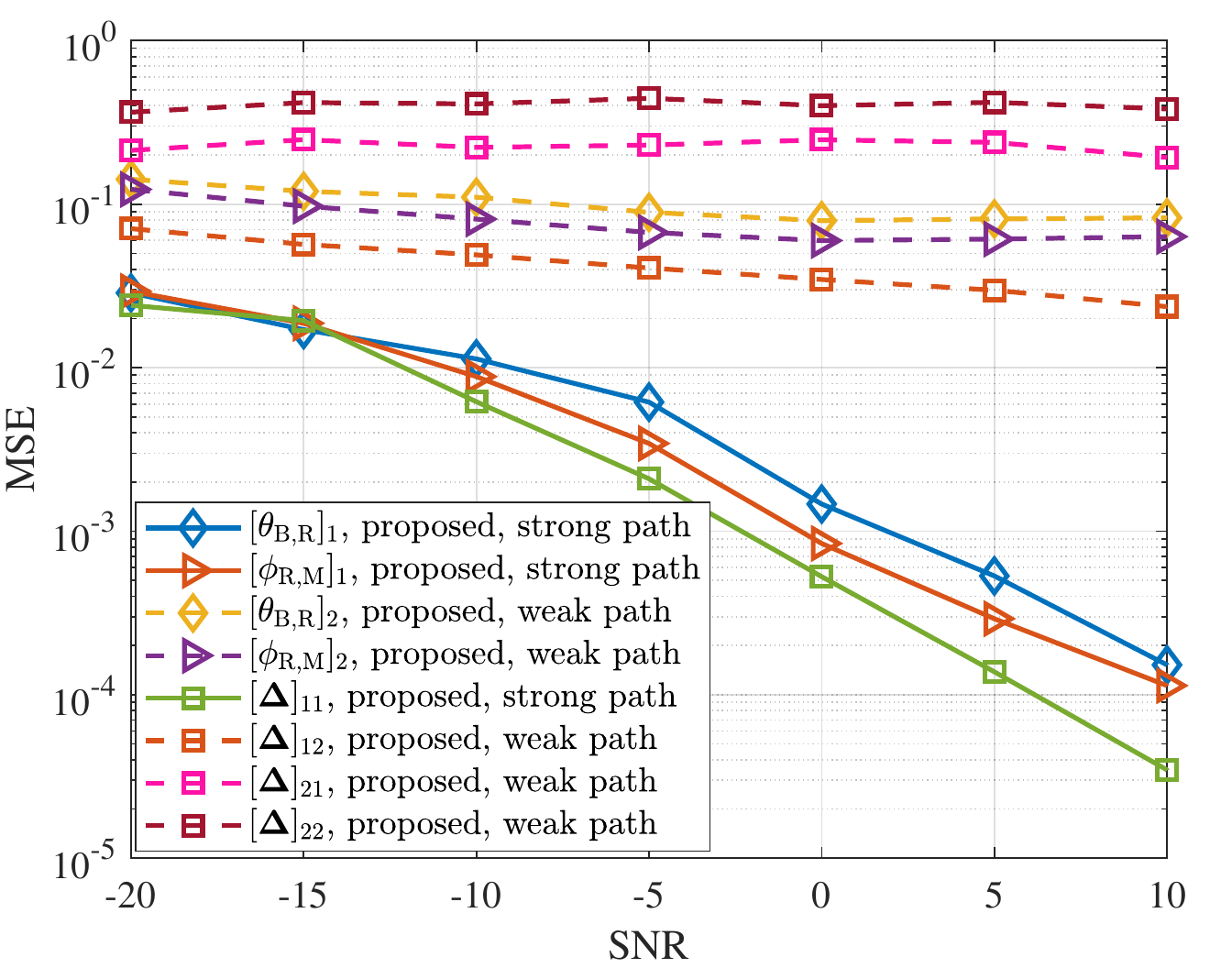}
	\caption{\tcb{Channel parameter estimation for inhomogeneous paths scenario from the path by path perspective. }}
	\label{parameter_estimate_inhomogeneous_paths}
\end{figure}

We continue to study the effect of path gain profile on the estimation performance. Unlike the homogenous paths with all the paths modelled as $\mathcal{CN}(0,1)$ in the previous subsection, we consider the scenario of inhomogenous paths with one path modelled as $\mathcal{CN}(0,1)$ and the remaining modelled as $\mathcal{CN}(0,0.01)$. On the average, 20 dB gap is considered regarding the average power of the strongest path vs.\ that of a weak path. The simulation parameters are set as $N_0 = M_0 = T= 10$ and $L_{\text{B,R}} = L_{\text{R,M}} =2$.  The simulation results on channel parameter estimation are provided in~Fig.~\ref{parameter_estimate_inhomogeneous_paths} path by path, where the prior information on the number of paths is assumed to be known precisely. The MSE performance of parameter estimation related to the strong paths  outperforms that related to the weak path(s).

\begin{figure}[t]
	\centering
	\includegraphics[width=.55\linewidth,trim={0 0 3 6},clip]{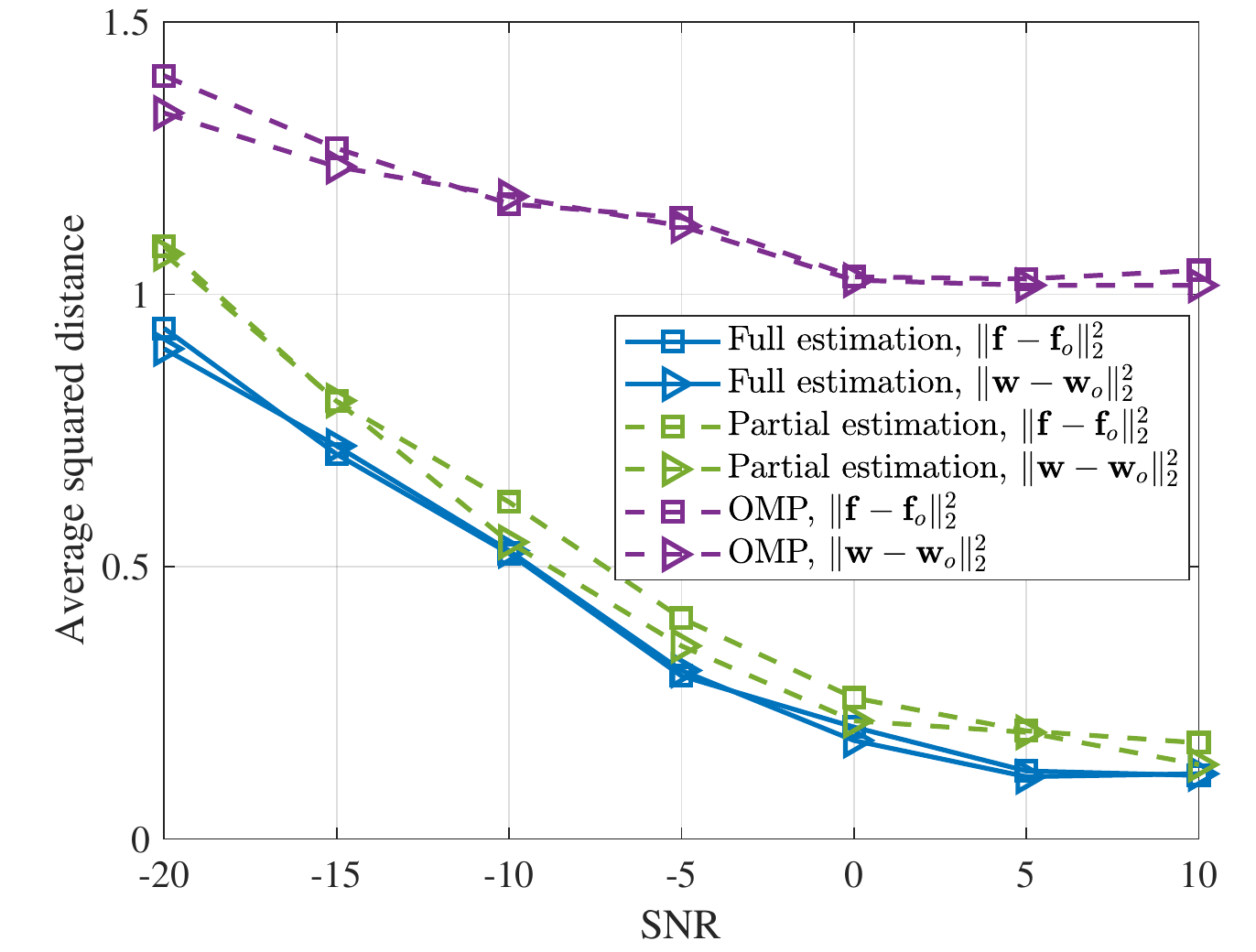}
	\caption{\tcb{Average squared distance between the designed beamformer/combiner and the optimal one for partial estimation vs.\ full estimation.} }
	\label{distance_of_bfs_combiners}
\end{figure}

We now study the ASD between the designed beamformer (combiner) in~Section~\ref{BF_at_BS_MS} and the optimal one, designed by assuming full CSI of the individual channels. We compare the performance with \emph{partial estimation}, 
where in Stage 2 only beams towards the strongest path are formed (leading to a reduced $T_t$). We also compare with the OMP-based two-stage approach. The performance is shown in Fig.~\ref{distance_of_bfs_combiners}. From the figure, we observe that the partial estimation can offer comparable performance compared to the that by full estimation in the inhomogeneous paths scenario, where only one path dominates in each individual channel. The performance of the proposed scheme significantly outperforms that of the OMP-based counterpart in terms of ASD. 

\begin{figure}[t]
	\centering
	\includegraphics[width=.55\linewidth,trim={0 0 3 6},clip]{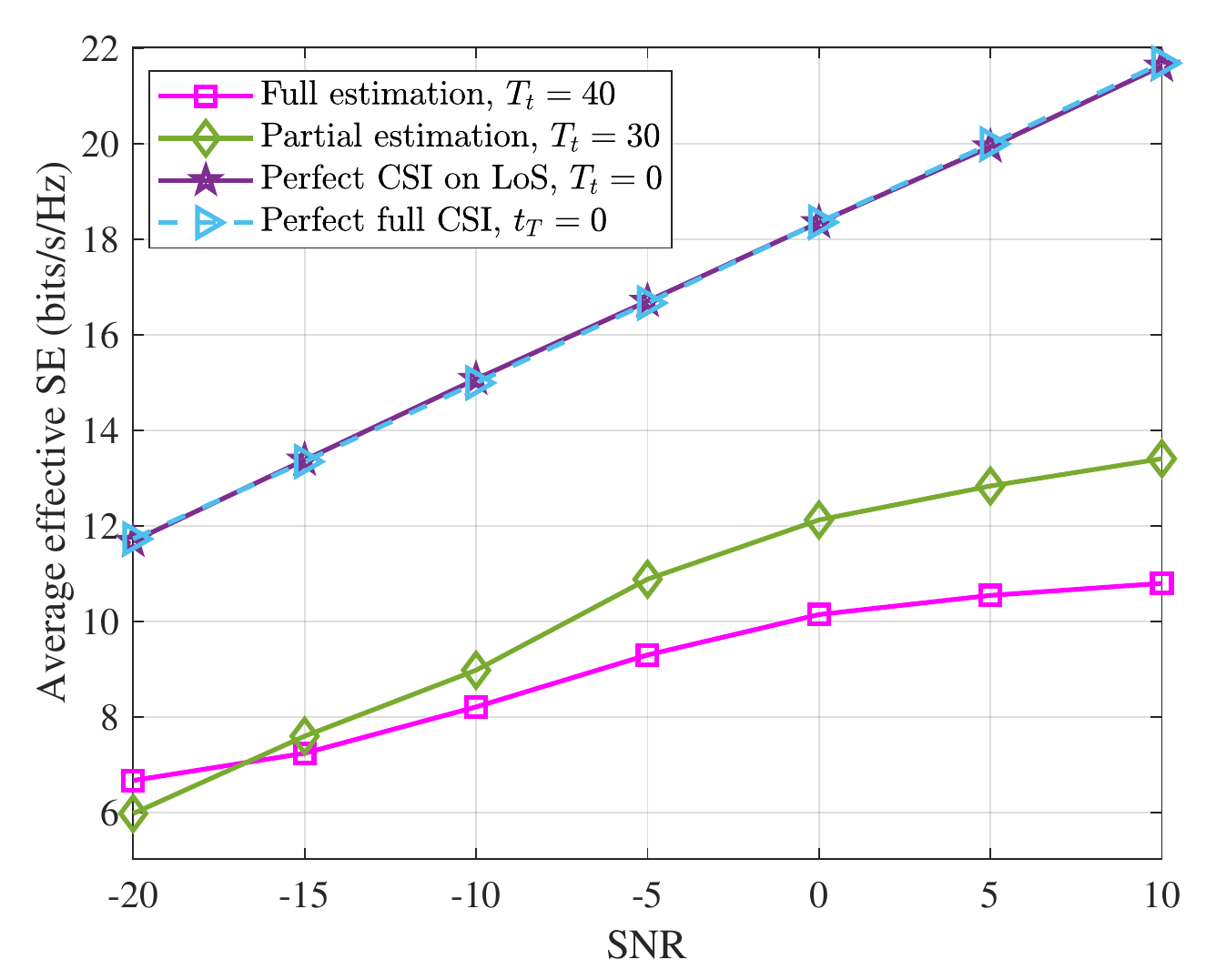}
	\caption{\tcb{Average effective SE bound for inhomogeneous paths scenario, partial estimation vs. full estimation.} }
	\label{average_SE_partial_full}
\end{figure}

The full estimation aiming at estimating all the channel parameters even brings some negative effect on the average effective SE bound, shown in Fig.~\ref{average_SE_partial_full}, compared to the partial estimation. This may result from the poor estimation of product of propagation path gains, related to weak paths, which in turn provides a bad design of RIS phase control matrix. An initial result on perfect CSI on the LoS (assuming that the strongest path is the LoS with path gain following $\mathcal{CN}(0,1)$) is obtained by aligning the beams towards the corresponding angles. As shown in Fig.~\ref{average_SE_partial_full}, knowing the LoS path (e.g., from the accurate location information) even brings some gains compared to the proposed scheme in the scenario of inhomogeneous paths, and offers similar performance with perfect full CSI case. This will attract great interests on application of location information (in practice imperfect) to the RIS-aided mmWave MIMO systems to boost the CE process and BF design.   

\section{Conclusions and Future Work}\label{section_conclusion_future_work}
We have studied the CE problem for the RIS-aided mmWave MIMO systems and proposed a two-stage atomic norm minimization problem, which can efficiently perform super-resolution channel parameter estimation. The power maximization criterion has been utilized to guide the design of phase control matrix at the RIS, followed by joint design of beamforming and combining vectors at the BS and MS based on the reconstructed composite channel. Simulation results have confirmed the advantages of the proposed scheme compared to the two-stage OMP approach in terms of MSE of angular parameter estimation and product of propagation of path gains estimation, average effective SE bound, and RIS gains in the homogeneous paths scenario. In the inhomogeneous paths scenario, we have evaluated the parameter estimation from the path by path perspective, where better performance can be achieved for the parameters related to the strong paths. The benefits brought by the availability of location information in the inhomogenous paths scenario has also been examined. 

Future studies can include the optimization of training and combining matrices during stage 1 sounding, 
optimization of the  regularization parameter to  bring a better trade-off between the data fitting (i.e., effect of noise term) and sparsity (i.e., prior information). 
In addition, the transmit powers during the entire sounding process can be optimized to bring better estimation performance. The prior information on the number of paths should be avoided to make the proposed scheme practical. Some preliminary results on the benefits brought by  location information on the RIS and MS are provided, and deserve to be explored in depth with a more realistic assumption on the location awareness.

\bibliographystyle{IEEEtran}
\bibliography{IEEEabrv,Ref_first_round}

\end{document}